\documentclass[a4paper, 12pt]{article}
\usepackage{amsfonts, amsthm, amsmath}
\usepackage[utf8]{inputenc}

\usepackage{amssymb}
\usepackage{graphicx}
\usepackage{scalerel,stackengine}
\usepackage{lscape}
\usepackage{xcolor}
\usepackage{cite}
\newtheorem{rem}{Remark}[section]
\newcommand\pig[1]{\scalerel*[5pt]{\big#1}{%
		\ensurestackMath{\addstackgap[1.5pt]{\big#1}}}}

\numberwithin{equation}{section}
\title{ {\bf
Superintegrable systems with spin and second-order (pseudo)tensor 
integrals of motion }}

\author{\vspace{1cm} \\
 {\bf \.{I}smet Yurdu\c{s}en}$^1$
        \thanks{E-mail address:
       yurdusen@hacettepe.edu.tr} \,,
      {\bf O. O\u{g}ulcan Tuncer}$^{1}$
        \thanks{E-mail address:
       otuncer@hacettepe.edu.tr} 
      {\,, and  \bf Pavel Winternitz}$^{2,3}$
       \thanks{E-mail address:
        wintern@crm.umontreal.ca}
     \\
   \\ $^1$Department of Mathematics, Hacettepe University,
                     \\ 06800 Beytepe, Ankara, Turkey
    \\
  \\ $^2$Centre de Recherches Math\'{e}matiques,  
    Universit\'{e} de Montr\'{e}al,\\ CP 6128, Succ. Centre-Ville, Montr\'{e}al, 
     Quebec H3C 3J7, Canada
  \\
   \\$^3$D\'{e}partement de Math\'{e}matiques et de Statistique, 
   Universit\'{e} de Montr\'{e}al,\\ CP 6128, Succ. Centre-Ville, Montr\'{e}al, 
  Quebec H3C 3J7, Canada}

\date{\today}
\newpage
\begin{document}
\setlength{\baselineskip}{24pt} 
\maketitle
\setlength{\baselineskip}{7mm}
\begin{abstract}
We investigate a quantum non-relativistic system describing the interaction of two particles with spin $\frac{1}{2}$ and spin 0, respectively. 
Assuming that the Hamiltonian is rotationally invariant and parity conserving we identify all such systems which allow additional (pseudo)tensor integrals of motion that are second order matrix polynomials in the momenta. Previously we found all the (pseudo)scalar and (axial)vector  integrals of motion. No non-obvious tensor integrals exist. However, nontrivial pseudo-tensor integrals do exist. Together with our earlier results we give a complete list of such superintegrable Hamiltonian systems allowing second-order integrals of motion.
\end{abstract}


\newpage
\section{Introduction}
\label{intro}
The purpose of this article is to report on a research program 
which investigates the superintegrability properties of systems involving 
particles with spin. This time the contribution is given by analyzing the second-order 
(pseudo)tensor integrals of motion. 
In this program we consider two non-relativistic 
particles with spin $s = 1/2$ and $s = 0$, respectively, which can be interpreted {\it e.g.} 
as a nucleon--pion interaction or an electron--$\alpha$ particle one. An earlier article 
\cite{DWY} was devoted to the investigation 
of the second-order (pseudo)scalar and (axial)vector integrals of motion 
for those Hamiltonian systems in a real three-dimensional Euclidean space. Hence, this work completes the 
full classification problem for those systems with second-order integrals of motion. 
The other articles in this program were devoted to the investigation of the same problem with first-order 
integrals of motion in $E_2$ and $E_3$ \cite{Winternitz.c,WYprooceeding,wy3} and second-order 
integrals of motion in $E_2$ \cite{yurdusen}. 

For a Hamiltonian system with $n$ degrees of freedom, integrability (in the Liouville sense, of course) corresponds to 
existing of $n$ independent commuting integrals of motion (including the Hamiltonian). If more than $n$ independent 
integrals exist, the system is called  superintegrable. 

The best known superintegrable systems are: Kepler, or Coulomb system ($V_0=\frac{\alpha}{r}$) and the
Harmonic oscillator ($V_0 = \omega r^2$). These are characterized by the fact that all finite classical
trajectories in these systems are periodic. Indeed, due to Bertrand's theorem \cite{Bertrand} these are the 
only two spherically symmetric potentials in which all bounded trajectories are closed. Releasing the restriction 
of spherical symmetry for the potential can lead to many new possibilities 
{\it e.g.} the anisotropic harmonic oscillator with rational ratio of frequencies \cite{Jauch}.

A systematic search for the properties of superintegrable systems was started quite some 
time ago \cite{Fris, Winternitz.b, Makarov}. Originally the approach concentrated on the natural Hamiltonians 
of type 
\begin{equation}
H=-\frac{1}{2} \Delta + V(\vec{r}\,)\,,
\label{introscalar}
\end{equation}
with integrals of motion that are second-order 
polynomials in the momenta and directly related with the multiseparability in two- and three-dimensional 
Euclidean spaces \cite{Fris, Evans.a, Evans.b, Miller.a}. This relationship between integrability and separability 
breaks down in other cases. For example, for natural Hamiltonians \eqref{introscalar}, the existence of third-order 
integrals does not lead to the separation of variables \cite{Drach, Gravel.a, Gravel.b}. Furthermore, 
if we consider velocity dependent potentials 
\begin{equation}
H=-\frac{1}{2} \Delta + V(\vec{r}\,)+(\vec{A}, \vec{p})\,, 
\label{introvector}
\end{equation}
then quadratic integrability no longer implies the separation of variables \cite{Dorizzi, Berube}. 

Second-order superintegrability has also been studied in 2- and 3-dimensional spaces of constant and nonconstant curvature 
\cite{Kalnins.a,Grosche1,Grosche2,Grosche3},\cite{Miller.b,Kalnins.b,Kalnins.c,Kalnins.d,Kalnins.e,Kalnins.f,Sheftel,Tempesta} and also in $n$ dimensions \cite{Rodriguez,Kalnins.h,Kalnins.i}. 

After the discovery of infinite families of classical and quantum systems with integrals of arbitrary order \cite{Tremblay.b}, 
the direction of the research has been shifted to {higher-order} integrability/superintegrability 
\cite{Tremblay.a,Marquette.b,Tremblay.c,Post,Kalnins.g,Kalnins.j,Kalnins.k,Quesne,Calzada,Marquette.c,Popper,Levesque,Chanu,MillerPostWinternitz:2013,Millerebook,PostWinternitz:2015}. Recently extended reviews have been published describing the current 
status of the subject \cite{MillerPostWinternitz:2013, Millerebook}.

More recently, exotic and standard potentials appearing in classical and quantum superintegrable systems 
have been studied both in Cartesian and polar coordinates 
\cite{MSW, AMJVPW2015,AMJVPWIY2018,AW,IanMproc,IanMproc2, EWY,Yurdusenconf, MPR, ELW}.

Previous studies of superintegrable systems with spin include \cite{Pronko.a, DHoker, Feher, Pronko.b, Nikitin.a, Nikitin.b, Nikitin.d, Nikitin.e, Nikitin.f}. 

The outline of the paper is as follows. In the next section, we introduce the formulation of the problem. Then, in section 3, 
we give the most general second-order tensor integrals of motion, obtain the determining equations and solve them in order to 
find all the second-order tensor integrals of motion. In section 4, we repeat the similar analysis for pseudo-tensor integrals of motion. 
Finally, in the last section we give a complete list of second-order integrals of motion and conclude about our results.  

\section{Formulation of the problem} 
\label{formulation}
We consider the following Hamiltonian
\begin{equation}
H=-\frac{\hbar^2}{2} \Delta + V_0(r) + V_1(r)\, (\vec{\sigma}, \vec{L})\,,
\label{intro1}
\end{equation}
in real three-dimensional Euclidean space $E_3$. Here $H$ is a matrix operator acting on a 
two-component spinor and we decompose it in terms of the $2 \times 2$ identity matrix $I$
and Pauli matrices (we drop the matrix $I$ whenever this does not cause confusion). We assume that the scalar potential $V_0(r)$ and the spin-orbital one $V_1(r)$ depend on the (scalar) distance $r$ only. The same system was already considered in \cite{DWY, wy3}. In \cite{DWY} the search for superintegrable systems was restricted to second-order (pseudo)scalar 
and (axial)vector  integrals of motion and in \cite{wy3} the search was restricted to first-order integrals.

Here we will investigate the second-order (pseudo)tensor integrals of motion. These are second-order matrix polynomials in the momenta. 
Our notations are the same as in \cite{wy3}, {\it i.e.}
\begin{equation}
 p_k = -i\hbar \partial_{x_k}\,, \qquad L_k = -i\hbar \epsilon_{klm}x_l\partial_{x_m}\,,
\end{equation}
are the linear and angular momentum respectively and
\begin{equation}
\sigma_1 = 
\left( \begin{array}{cc}
0 & 1 \\
1 & 0 \end{array} \right)\,, \qquad
\sigma_2 = 
\left( \begin{array}{cc}
0 & -i \\
i & 0 \end{array} \right)\,, \qquad
\sigma_3 = 
\left( \begin{array}{cc}
1 & 0 \\
0 & -1 \end{array} \right)\,,
\end{equation}
are the Pauli matrices.

The system \eqref{intro1} is
integrable by construction. Since $V_0(r)$ and $V_1(r)$ depend only on the distance $r$ and $(\vec{\sigma},\vec{L})$ is a scalar, the Hamiltonian $H$ commutes with the total angular momentum $\vec{\mathcal{J}}$ which is defined by
\begin{equation}
	\vec{\mathcal{J}}=\vec{L}+\dfrac{\hbar}{2}\vec{\sigma}.
\end{equation}

To obtain nontrivial results, we impose from the beginning that the spin-orbital interaction be present ($V_1 \neq 0$). We also recall a result from \cite{wy3}, namely for 
\begin{equation}\label{gauge1}
	V_1(r) = \frac{\hbar}{r^2}\,, \qquad V_0(r) \quad \text{arbitrary}\,,
\end{equation}
the Hamiltonian (\ref{intro1}) allows 2 first-order axial vector integrals of motion. They are $\vec{\mathcal{J}}$ and
\begin{equation}
	\label{Saxial}
	\vec{S}=-\frac{\hbar}{2} \vec{\sigma}+\hbar \frac{\vec{x}}{r^2} (\vec{x},\vec{\sigma})\,.
\end{equation}
For
\begin{equation}\label{gauge2}
	V_1(r) = \frac{\hbar}{r^2}\,, \qquad V_0(r) =\frac{\hbar^2}{r^2}\,,
\end{equation}
it allows 2 first-order axial vector integrals and 1 first-order vector integral. They are $\vec{\mathcal{J}}$, $\vec{S}$ and 
\begin{equation}
	\label{Piaxial}
	\vec{\Pi} = \vec{p}-\frac{\hbar}{r^2} (\vec{x}\wedge\vec{\sigma})\,.
\end{equation}

These two systems are first-order superintegrable and the term $V_1(r) = \frac{\hbar}{r^2}$ can be induced from a Hamiltonian with $V_1(r) = 0$ by a  gauge transformation \cite{wy3}. For the gauge induced potentials given in \eqref{gauge1}, the integrals of motion are either just the gauge transforms of the terms obtained from $\vec{L}$, $\vec{\sigma}$, or else leave in the enveloping algebra of a direct sum of the algebra $o(3)$ with itself ${o}(3)\oplus {o}(3) = \{\vec{\mathcal{J}}-\vec{S}\}\oplus \{\vec{S}\}$. For the gauge induced potentials given in \eqref{gauge2}, such integrals of motion are either the gauge transforms of the terms obtained from $\vec{L}$, $\vec{p}$, $\vec{\sigma}$, or else leave in the enveloping algebra of a direct sum of the Euclidean Lie algebra $e(3)$ with the algebra $o(3)$: $e(3)\oplus {o}(3) = \{\vec{\mathcal{J}}-\vec{S}, \vec{\Pi}\}\oplus \{\vec{S}\}$. (For details; see \cite{wy3}.) Hence, in the subsequent  analysis we exclude the cases when the spin-orbital potential is a gauged induced one. However, we will present those pseudo(tensor) integrals of motion corresponding to gauge induced potentials in the appendix in order to explicitly verify that they reduce to the well-known results in the absence of the spin-orbit interaction. 

In sections 3 and 4, we search for second-order (pseudo)tensor integrals commuting with the Hamiltonian \eqref{intro1}. Such commutations yield the so-called determining equations which will then be solved to find the integrals of motion. The Hamiltonian \eqref{intro1} is invariant under rotations and reflections in $E_3$ which, however, can transform the integrals of motion into new invariants. Thus, instead of considering the solution of the whole set of determining equations, it is reasonable to simplify the problem by classifying the integrals of motion into irreducible $O(3)$ multiplets.

At our disposal are two vectors $\vec{x}$ and $\vec{p}$ and one pseudo-vector $\vec{\sigma}$. The integrals we are
considering can involve at most second-order powers of $\vec{p}$ and first-order powers of $\vec{\sigma}$, but arbitrary powers of $\vec{x}$. 

We shall construct symmetric two-component tensors and pseudo-tensors in the space:
\begin{equation}\label{sp}
\{\{\vec{x}\}^n\times\vec{p}\times\vec{\sigma}\}.
\end{equation}
The quantities $\vec{x}$, $\vec{p}$ and $\vec{\sigma}$ allow us to define six independent `directions' in the direct product of the Euclidean space and the spin one, namely 
\[ \{ \vec{x},\vec{p},\vec{L}=\vec{x}\wedge\vec{p},\vec{\sigma},\vec{\sigma}\wedge\vec{x},\vec{\sigma}\wedge\vec{p} \}, \]
and any $O(3)$ tensor can be expressed in terms of these. The positive integer $n$ in \eqref{sp} is arbitrary and any scalar in $\vec{x}$ space will be written as $f(r)$, where $f$ is an arbitrary function of $r=\sqrt{x^2+y^2+z^2}$. Since $\vec{p}$ figures at most quadratically and $\vec{\sigma}$ at most linearly, we can form 28 two-component symmetric tensors and 26 symmetric pseudo-tensors. In the following sections we will separately and explicitly give the form of these integrals of motion.



\section{Tensor integrals of motion} 
Two index tensors are expressed as follows:
\begin{align}\label{tensor}
&T_1^{ij}=x^ix^j,\quad T_2^{ij}=(\vec{x},\vec{p})x^ix^j,\quad T_3^{ij}=(\vec{\sigma},\vec{L})x^ix^j,\quad T_4^{ij}=x^ip^j+p^ix^j, \nonumber\\ & T_5^{ij}=x^i(\vec{x}\wedge\vec{\sigma})^j+(\vec{x}\wedge\vec{\sigma})^ix^j,\quad T_6^{ij}=(\vec{x},\vec{p})(x^i(\vec{x}\wedge\vec{\sigma})^j+(\vec{x}\wedge\vec{\sigma})^ix^j),\quad T_7^{ij}=L^i\sigma^j+\sigma^i L^j, \nonumber\\& T_8^{ij}=p^i(\vec{x}\wedge\vec{\sigma})^j+(\vec{x}\wedge\vec{\sigma})^ip^j, \quad T_{9}^{ij}={\vec{p}}^{\:2}x^ix^j,\quad  T_{10}^{ij}=\vec{L}^2x^ix^j,\quad 
T_{11}^{ij}=(\vec{x},\vec{p})(\vec{\sigma},\vec{L})x^ix^j, \nonumber\\
& T_{12}^{ij}=L^iL^j+L^jL^i, \quad T_{13}^{ij}=(\vec{\sigma},\vec{L})(x^ip^j+p^ix^j),\quad T_{14}^{ij}=\vec{p}^{\:2} (x^i(\vec{x}\wedge\vec{\sigma})^j+(\vec{x}\wedge\vec{\sigma})^ix^j),\nonumber\\ & T_{15}^{ij}=\vec{L}^2(x^i(\vec{x}\wedge\vec{\sigma})^j+(\vec{x}\wedge\vec{\sigma})^ix^j),\quad T_{16}^{ij}=(\vec{x},\vec{p})(x^i(\vec{p}\wedge \vec{\sigma})^j+(\vec{p}\wedge \vec{\sigma})^ix^j),\quad T_{17}^{ij}=p^ip^j,\nonumber\\ & T_{18}^{ij}=(\vec{x},\vec{p})(p^i(\vec{x}\wedge\vec{\sigma})^j+(\vec{x}\wedge\vec{\sigma})^ip^j),\quad T_{19}^{ij}=p^i(\vec{p}\wedge\vec{\sigma})^j+(\vec{p}\wedge\vec{\sigma})^ip^j,\nonumber\\
& T_{20}^{ij}=x^i(\vec{p}\wedge\vec{\sigma})^j+(\vec{p}\wedge\vec{\sigma})^ix^j,\quad T_{21}^{ij}=(\vec{x},\vec{p})(L^i\sigma^j+\sigma^i L^j),\quad
T_{22}^{ij}=(\vec{x},\vec{p})(x^ip^j+p^ix^j),\nonumber\\ & T_{23}^{ij}=(\vec{x},\vec{p})^2x^ix^j,\quad T_{24}^{ij}=(\vec{x},\vec{p})^2(x^i(\vec{x}\wedge\vec{\sigma})^j+(\vec{x}\wedge\vec{\sigma})^ix^j),\quad T_{25}^{ij}=(\vec{\sigma},\vec{x})(x^iL^j+L^ix^j),\nonumber\\
& T_{26}^{ij}=(\vec{x},\vec{p})(\vec{\sigma},\vec{x})(x^iL^j+L^ix^j),\quad T_{27}^{ij}=(\vec{\sigma},\vec{x})(p^iL^j+L^ip^j),\quad T_{28}^{ij}=(\vec{\sigma},\vec{p})(x^iL^j+L^ix^j).
\end{align}
Each of the quantities in \eqref{tensor} can be multiplied by a scalar $f(r)$ without changing its properties under rotations or reflections. 

It can be shown that we have three linear relations
\begin{align}\label{eq0}
T_{20}^{ij}=-T_7^{ij}+T_8^{ij},\qquad
T_{21}^{ij}=-T_{16}^{ij}+T_{18}^{ij},\qquad 
T_{28}^{ij}=T_{14}^{ij}-T_{16}^{ij}+T_{13}^{ij}\,,
\end{align}
and even though all the above tensors but $T_{20}^{ij},T_{21}^{ij}$ and $T_{28}^{ij}$ are linearly independent, higher-order polynomial relations between them exist. 
Indeed, we have the following syzygies:
\begin{align}
&T_{22}^{ij}=\frac{1}{2}(\vec{r}^{\:2}T_{17}^{ij}+T_{12}^{ij}+T_{9}^{ij}),\nonumber\\
&T_{23}^{ij}=\vec{r}^{\:2}T_{9}^{ij}-T_{10}^{ij},\nonumber\\
&T_{24}^{ij}=\vec{r}^{\:2}T_{14}^{ij}-T_{15}^{ij},\nonumber\\
&T_{25}^{ij}=-\vec{r}^{\:2}T_7^{ij}+T_6^{ij}+T_{3}^{ij},\nonumber\\
&T_{26}^{ij}=-\vec{r}^{\:2}T_{16}^{ij}+T_{24}^{ij}+T_{11}^{ij},\nonumber\\
&T_{27}^{ij}=-\vec{r}^{\:2}T_{19}^{ij}+T_{18}^{ij}+T_{13}^{ij}\label{eq1}. 
\end{align}
We use relations \eqref{eq0} and \eqref{eq1} to remove the left-hand sides
of these equations from the analysis completely.
\subsection{The commutativity condition and determining equations}
In this subsection, we take the linear combinations of all the two-component tensors 
and then fully symmetrize them in order to obtain the determining equations from the commutativity condition.

Let us take the linear combination of the independent tensors given in \eqref{tensor} as 
\[X_T^{ij}=\sum\limits_{k=1}^{19}f_k(r)T_{k}^{ij}, \]
which can be fully symmetrized as decribed in \cite{DWY}. In the commutativity relation $[H,X_T^{ij}]=0$ it is enough to consider only two indices, say $i=1, j=2$ since the others then necessarily commute due to the rotations. In the analysis we consider the full symmetric form of 
$X_T^{12}$ which, however, is rather long to be presented here. From the requirement $[H,X_T^{ij}]=0$, we obtain the determining equations.

The determining equations obtained by equating the coefficients of third-order terms to zero in the commutativity equation give us 
\begin{align}
& f_{10}=0,\quad f_{9}=0, \quad f_{12}=c_1,\quad
f_{17}=c_2,\quad  f_{11}=-f_{15}, \label{eq4} \medskip \\ & f_{16}=-f_{14},\quad
f_{19}=-r^2 f_{18}+c_4, \quad f_{13}=r^2 f_{15}+f_{14}+f_{18},\quad f_{14}=c_3,\label{eq5} \medskip \\
&2rf_{15}V_1+\hbar f_{15}'=0, \label{eq8}\\
&2c_3V_1-\hbar (f_{15}+ r f_{15}')=0, \label{eq9}\\
&2r(c_3+r^2f_{15}-f_{18})V_1-\hbar f_{18}'=0, \label{eq10}\\
&\hbar f_{18}+2c_4V_1+\hbar(c_3+rf_{18}')=0, \label{eq11}
\end{align}
where $c_i\,(i=1,2,3,4)$ are integration constants.

Introducing the relations given in
\eqref{eq4} and \eqref{eq5} into the determining equations obtained by equating the coefficients of the second-order
terms to zero in the commutativity equation, we get \allowdisplaybreaks
\begin{align}
& f_{3}=f_6+c_5, \quad f_7=r^2f_6+c_6,  \label{eqq10}\\
&rf_{15}V_1-f_2'=0,\label{eqq11}\\
&4rf_2+r(c_3+2r^2f_{15}+f_{18})V_1+f_4'-2c_4V_1'=0,\label{eqq12}\\
&2c_5\hbar r+2rf_6(-\hbar+r^2V_1)-\hbar(f_8'+2c_2V_1')=0,\label{eqq13}\\
&2 r f_{6} V_1+{\hbar} f_{6}'=0,\label{eqq14}\\
&-3\hbar rf_6-2r(c_6-c_1\hbar+f_8)V_1-\hbar f_8'=0,\label{eqq15}\\
&2rf_2+r(-3c_3-2r^2f_{15}+3f_{18})V_1+f_4'=0,\label{eqq16}\\
&f_4+(3c_4+c_3r^2+r^4f_{15}-2r^2f_{18})V_1+c_4rV_1'=0,\label{eqq17}\\
&2c_5\hbar r+\hbar f_6+2r(c_6-c_1\hbar+r^2f_6+f_8)V_1-2c_2\hbar V_1'=0  \label{eqqq18}\\
&-2c_6V_1+\hbar(f_6+2c_1V_1+rf_6')=0,\label{eqq18}\\
&f_8-2c_2V_1+r(3rf_6+r^2f_6'+f_8')=0,\label{eqq19}
\end{align}
where, again, $c_5$ and $c_6$ are integration constants.

Now, introducing the relations given in
\eqref{eq4}, \eqref{eq5} and \eqref{eqq10} into the determining equations obtained by equating the coefficients of the first-order
terms to zero in the commutativity equation yields \allowdisplaybreaks
\begin{align}
&r(2c_5-f_6)V_1-f_1'+(r^2f_6+f_8)V_1'=0, \label{eqq21}\\
& 4c_2V_0'-2r\pig(2f_1+\big(2r^2(c_5+f_6)+3f_8\big)V_1+r(r^2f_6+f_8)V_1'\pig) =0,  \label{eqq22}\\
&12\hbar rf_{15}V_1+8\hbar^2f_{15}'-2\hbar f_{18}'V_1+4f_5'-4c_3\hbar V_1'-4\hbar f_{18}V_1'+4f_4V_1'+4rf_2(-2V_1+rV_1')\nonumber\\ &\qquad\qquad\qquad+\hbar^2 rf_{15}'' =0,  \label{eqq23}\\
& -4rf_5+6\hbar rf_{18}V_1+4rf_4V_1-2\hbar rf_{15}(\hbar-5r^2V_1)+8\hbar^2r^2f_{15}'+10\hbar^2f_{18}'-2\hbar r^2f_{18}'V_1 \nonumber\\ &\qquad\qquad\qquad +4c_4V_0'+4c_4\hbar V_1'+\hbar^2 r^3f_{15}''+2\hbar^2 rf_{18}'' =0.\label{eqq24}
\end{align}

Finally, introducing the relations given in
\eqref{eq4}, \eqref{eq5} and \eqref{eqq10} into the determining equations obtained by equating the coefficients of the zeroth-order
terms to zero in the commutativity equation, we obtain \allowdisplaybreaks
\begin{align}
& -2c_2V_0'+r\pig( -r\big(6f_1'+3V_1(-2c_5r+3rf_6+r^2f_6'+f_8')+rf_1'' \big)+2c_2V_0'' \pig)=0,\label{eqq25}\\
& 12r^2(-r f_5 V_1+3\hbar f_2')+4r^2(r^2f_2+f_4)V_0'+\hbar^2\pig(-4f_4'+r\big(4f_4''+r(13rf_2''+r^2f_2^{(3)}+f_4^{(3)})\big)\pig) =0,\label{eqq26}\\
&  10\hbar r^3f_2V_1-2r^3f_5V_1+\hbar\pig(-2c_4V_0'+r\big( r(2V_1(r^2f_2'+f_4')-6f_5'+(c_3+f_{18})V_0'-rf_5'')\nonumber\\ &\qquad\qquad\qquad+2c_4V_0''\big) \pig)=0\label{eqq27},
\\
&  24\hbar^2r^2f_6'+2r^2V_1\big(4rf_1+\hbar(-2c_5r+3rf_6+r^2f_6'+f_8') \big)+4r^2(r^2f_6+f_8)V_0'  \nonumber\\ &\qquad\qquad\qquad 
+\hbar^2 \pig(-4f_8'+r\big( 4f_8''+r(11rf_6''+r^2f_6^{(3)}+f_8^{(3)}) \big)  \pig) =0.\label{eqq28}
\end{align}
We notice that there are four types of determining equations:

\begin{enumerate}
	\item  Those that are independent of the potentials $V_1$ and $V_0$, namely \eqref{eq4}, \eqref{eq5} and \eqref{eqq10}. We input these relations into the other determining	equations. This eliminates some of the equations involving the potential
	and we have presented only the remaining ones.
	\item  Equations involving only $V_1$ but not its derivative \eqref{eq8}-\eqref{eq11}, \eqref{eqq11}, \eqref{eqq14}-\eqref{eqq16}, \eqref{eqq18} and \eqref{eqq19}.
	\item Those involving $V_1$ and $V'_1$ which are \eqref{eqq12}, \eqref{eqq13}, \eqref{eqq17}, \eqref{eqq21} and \eqref{eqq23}.
	\item Those involving  $V_0$, $V_1$ and their derivatives.
\end{enumerate}


\subsection{Solutions of the determining equations}  
In this subsection, we find the solutions of the determining equations coming from the tensors and then obtain the tensor integrals of motion. We present a complete classification by following the above four classes of determining equations sequentially. As indicated before, the first type of determining equations had already been introduced into the others to eliminate or simplify them. So, we start the analysis with ten determining equations of second type. These equations can be viewed as linear algebraic equations for $V_1$. They must all be satisfied simultaneously. Thus they must all be multiples of just one equation. Its solution determines $V_1$ uniquely (though there may be several parametric versions of it depending on the coefficients $f_i$, $c_a$). 

Let us first consider \eqref{eq8}. Assuming $f_{15}\neq 0$, we obtain 
\begin{equation}\label{eqlast1}
	V_1 = -\dfrac{\hbar f_{15}'}{2r f_{15}}.
\end{equation}
This equation can be introduced into \eqref{eq9} to give
\begin{equation}\label{eqlast2-1}
	f_{15}=\dfrac{-c_3+\epsilon\sqrt{{c_3}^2+\gamma_1 r^2}}{r^2},
\end{equation}
where $c_3$ and $\gamma_1$ are real constants. First, it would be better to see what happens when $c_3=0$. The reason for this will become clear shortly. In this case, we immediately find
\begin{equation}
	f_{15}=\dfrac{\alpha}{r},
\end{equation}
where $\alpha$ is an integration constant. Introducing this into \eqref{eq8}, we find
\begin{equation}
	2\alpha V_1-\dfrac{\alpha \hbar}{r^2}=0,
\end{equation}
which gives us two options, either $\alpha=0$ or $V_1=\dfrac{\hbar}{2r^2}$. But, the first one yields $f_{15}=0$ which is undesirable due to the assumption. Then, let us take $V_1=\dfrac{\hbar}{2r^2}$. If this potential is considered in \eqref{eq10} and \eqref{eq11}, one can easily get $c_3=0$ and $\alpha=0$, which again yields $f_{15}=0$. So, we need to take $c_3\neq 0$ which lead to \eqref{eqlast2-1} to be rewritten as
\begin{equation}\label{eqlast2}
	f_{15}=-\dfrac{c_3(1+\epsilon\sqrt{1+\beta r^2})}{r^2},
\end{equation}
where $\epsilon^2=1$ and $\beta$ is a real constant. From \eqref{eqlast1} and \eqref{eqlast2} we get the following potential:
\begin{equation}\label{eqlast3}
	V_1=\dfrac{\hbar}{2r^2}\left(1+\dfrac{\epsilon}{\sqrt{1+\beta r^2}}\right).
\end{equation}
It is obvious that $V_1=\dfrac{\hbar}{r^2}$ and $V_1=\dfrac{\hbar}{2r^2}$ are special solutions for $(\epsilon,\beta)=(1,0)$ and $(1,\infty)$, respectively. 
The first special potential is a gauge induced one and had already been thoroughly considered in \cite{wy3}.  Hence, we exclude the analysis of this option which means that we will assume $\beta\neq 0$. 

By considering the above relations \eqref{eqlast2} and \eqref{eqlast3} in the equations \eqref{eq10} and \eqref{eq11}, one can obtain
\begin{equation}
	f_{18}=\dfrac{-c_3  r^2(1+\beta r^2)+c_4(1+\epsilon\sqrt{1+\beta r^2})}{r^2}.
\end{equation}
Substituting this back into \eqref{eq11} yields $\beta=0$ or $c_3=0$. The case $\beta=0$ gives either $V_1=0$ or $V_1=\dfrac{\hbar}{r^2}$ both of which are excluded.  On the other hand, from \eqref{eqlast2} the case $c_3=0$ means that $f_{15}=0$ which contradicts with the assumption $f_{15}\neq 0$.

The above discussion states that $f_{15}$ must vanish. So, \eqref{eq9} directly yields $c_3=0$. Keeping in mind these two facts, let us continue the analysis by considering \eqref{eq10}. Assume that $f_{18}\neq 0$. Then, one can easily get
\begin{equation}
		V_1 = -\dfrac{\hbar f_{18}'}{2r f_{18}} ,
\end{equation}
and introducing this into \eqref{eq11} we obtain
\begin{equation}\label{eqlast4-1}
	f_{18}=\dfrac{c_4+\epsilon\sqrt{{c_4}^2+\gamma_2 r^2}}{r^2},
\end{equation}
where $c_4$ and $\gamma_2$ are real constants. Similar to the above, it can be easily checked that the case $c_4=0$ will eventually cause $f_{18}=0$. Hence, we take $c_4\neq 0$ which allows us to rewrite \eqref{eqlast4-1} as
\begin{equation}\label{eqlast4}
	f_{18}=\dfrac{c_4(1+\epsilon\sqrt{1+\beta r^2})}{r^2},\quad \epsilon^2=1.
\end{equation}
Then, from the last two relations (3.33) and (3.34) we again have the potential
\begin{equation}\label{eqlast5}
	V_1=\dfrac{\hbar}{2r^2}\left(1+\dfrac{\epsilon}{\sqrt{1+\beta r^2}}\right).
\end{equation}
Then, by introducing these into \eqref{eqq14}, \eqref{eqq15}, \eqref{eqq18} and \eqref{eqq19} one can find
\begin{align}\label{eqlast6}
	&f_6=\dfrac{(c_1\hbar-c_6)(1+\epsilon\sqrt{1+\beta r^2})}{r^2},\\
	&f_8=(c_6-c_1\hbar)(1+\beta r^2+\epsilon\sqrt{1+\beta r^2})-\dfrac{\hbar c_2(1+\epsilon\sqrt{1+\beta r^2})}{r^2}.
\end{align}
Substituting these relations into \eqref{eqq15} gives $\beta(c_6-c_1 \hbar)=0$. Hence, we directly have $c_6=c_1\hbar$ since the case $\beta=0$ is excluded. 

The remaining two determining equations of second type, namely \eqref{eqq11} and \eqref{eqq16}, give
\begin{equation}\label{eqlast7}
	f_2=d_1,\quad f_4=-d_1r^2+\dfrac{3c_4\hbar(1+\epsilon\sqrt{1+\beta r^2})}{2r^2}+d_2,
\end{equation}
where $d_1$ and $d_2$ are real constants. Then, we see that for the relations \eqref{eqlast4}, \eqref{eqlast6}, \eqref{eqlast7} and the potential given in \eqref{eqlast5} all of the determining equations of second type are satisfied. Now, we need to continue the analysis with the determining equations of third type. However, one can easily check that introducing these relations into \eqref{eqq12} immediately yields either $(d_1,\beta)=(0,0)$ or $(d_1,c_4)=(0,0)$. We already know that the first case is excluded. The second case gives us a contradiction. This means that we do not need to examine the case $f_{18}\neq 0$ further. 

Hence, $f_{18}$ must vanish too, which immediately gives us $c_4=0$. We continue the analysis with the relations $f_{15}=0$, $c_3=0$, $f_{18}=0$ and $c_4=0$. By introducing these relations into \eqref{eqq11} and \eqref{eqq16} we find
\begin{equation}
	f_2=d_1,\quad f_4=-d_1r^2+d_2,
\end{equation}
where $d_1$ and $d_2$ are integration constants. 

Assuming $f_6\neq 0$ in \eqref{eqq14} we obtain
\begin{equation}
	V_1 = -\dfrac{\hbar f_{6}'}{2r f_{6}} ,
\end{equation}
and introducing this into \eqref{eqq18} yields 
\begin{equation}\label{eqlastnew}
	f_{6}=\dfrac{-(c_6-c_1\hbar)+\epsilon\sqrt{(c_6-c_1\hbar)^2+\gamma_3 r^2}}{r^2},
\end{equation}
where $c_1$, $c_6$ and $\gamma_3$ are real constants. From the remaining determining equations, one can see that the special case when $c_6=c_1 \hbar$ will give rise to $f_6=0$. Therefore, we need to assume $c_6\neq c_1\hbar$ which gives 
\begin{equation} 
	f_{6}=\dfrac{(c_1\hbar-c_6)(1+\epsilon\sqrt{1+\beta r^2})}{r^2},\quad \epsilon^2=1.
\end{equation}
Then, the last two relations once again give the potential
\begin{equation} 
	V_1=\dfrac{\hbar}{2r^2}\left(1+\dfrac{\epsilon}{\sqrt{1+\beta r^2}}\right).
\end{equation}
One can easily check that substituting these relations into \eqref{eqq15} and \eqref{eqq18} yields $(c_6-c_1 \hbar)\beta=0$. Thus, we have $c_6=c_1\hbar$ due to the fact that $\beta\neq 0$. However, this contradicts with the assumption $f_6\neq 0$. Therefore, we need to take $f_6=0$ which directly yields $c_6=c_1\hbar$ by \eqref{eqq18}. 

The above discussion implies that two determining equations of second type remain:
\begin{equation}
	2rf_8V_1+\hbar f_8'=0,\quad f_8+rf_8'-2c_2V_1=0.
\end{equation}
Let us first assume $f_8\neq 0$. In this case, it is rather easy to see that $c_2$ cannot vanish for this case. So, similar steps to the previous cases yield the relation
\begin{equation} 
	f_{8}=-\dfrac{c_2\hbar(1+\epsilon\sqrt{1+\beta r^2})}{r^2},\quad \epsilon^2=1, 
\end{equation}
and the potential
\begin{equation} 
	V_1=\dfrac{\hbar}{2r^2}\left(1+\dfrac{\epsilon}{\sqrt{1+\beta r^2}}\right).
\end{equation}
So, all of the determining equations of second type are satisfied. But, by introducing these relations into \eqref{eqq13} we see that either $(c_5,\beta)=(0,0)$ or $(c_5,c_2)=(0,0)$. Again, the first case is excluded and the second one causes a contradiction.

Eventually, by solving the determining equations of second type we just find the following relations for an arbitrary spin-orbital potential:
\begin{equation}\label{eqremain}
	f_{15}=0,\quad f_{18}=0,\quad f_6=0,\quad f_8=0,\quad c_3=0,\quad c_4=0,\quad c_6=c_1\hbar, \quad c_2=0. 
\end{equation}
Let us examine the determining equations of third type. Introducing the relations in \eqref{eqremain} into the determining equations \eqref{eqq12}, \eqref{eqq13}, \eqref{eqq17}, \eqref{eqq21} and \eqref{eqq23}, one can easily obtain
\begin{equation}\label{eqremain2}
	f_2=0,\quad f_4=0,\quad f_1=k_1,\quad c_5=0,
\end{equation}
where $k_1$ is an integration constant. By these relations, all of the determining equations of third type are satisfied. Considering \eqref{eqremain} and \eqref{eqremain2} in the remaining equations, one can see that $f_1=0$ and those equations are satisfied for arbitrary scalar potential $V_0$. Hence, we have just two functions $f_{12}$ and $f_{7}$ that do not vanish. 

Finally, all the determining equations are satisfied for arbitrary potentials $V_0=V_0(r)$ and $V_1=V_1(r)$. Having left with just one arbitrary constant $c_1$, we have the following trivial integral of motion 
\begin{equation}\label{intmot}
	\mathcal{T}_1^{\,ij}=\hbar (L^i\sigma^j+\sigma^{i}L^j)+L^iL^j+L^jL^i=\{\mathcal{J}_i,\mathcal{J}_j\}-\delta_{ij}\dfrac{\hbar}{2},
\end{equation}
where $\{\cdot,\cdot\}$ stands for the anti-commutator.

\section{Pseudo-tensor integrals of motion} 
Two index pseudo-tensors are expressed as follows:
\begin{align}\label{pseudo}
&Y_1^{ij}=x^i\sigma^j+\sigma^ix^j,\quad Y_2^{ij}=(\vec{x},\vec{p})(x^i\sigma^j+\sigma^ix^j),\quad Y_3^{ij}=p^i\sigma^j+\sigma^ip^j,\quad Y_4^{ij}=(\vec{\sigma},\vec{x})x^ix^j, \nonumber\\ & Y_5^{ij}=(\vec{\sigma},\vec{p})x^ix^j,\quad Y_6^{ij}=(\vec{x},\vec{p})(\vec{\sigma},\vec{x})x^ix^j,\quad Y_7^{ij}=(\vec{\sigma},\vec{x})(x^ip^j+p^ix^j),\nonumber\\& Y_8^{ij}=x^iL^j+L^ix^j, \quad Y_{9}^{ij}={\vec{p}}^{\:2}(x^i\sigma^j+\sigma^ix^j),\quad  Y_{10}^{ij}=\vec{L}^2(x^i\sigma^j+\sigma^ix^j),\nonumber\\
& 
Y_{11}^{ij}=(\vec{x},\vec{p})(p^i\sigma^j+\sigma^ip^j), \quad Y_{12}^{ij}=(\vec{x},\vec{p})((\vec{x}\wedge\vec{\sigma})^iL^j+L^i(\vec{x}\wedge\vec{\sigma})^j), \nonumber\\ & Y_{13}^{ij}=(\vec{p}\wedge\vec{\sigma})^iL^j+L^i(\vec{p}\wedge\vec{\sigma})^j,\quad Y_{14}^{ij}=(\vec{\sigma},\vec{x})\vec{L}^{\:2}x^ix^j,\quad Y_{15}^{ij}=(\vec{x},\vec{p})(\vec{\sigma},\vec{x})(x^ip^j+p^ix^j),\nonumber\\ & Y_{16}^{ij}=(\vec{\sigma},\vec{x})p^ip^j,\quad Y_{17}^{ij}=(\vec{x},\vec{p})(x^iL^j+L^ix^j),\quad Y_{18}^{ij}=(\vec{\sigma},\vec{L})(x^iL^j+L^ix^j),\nonumber\\
& Y_{19}^{ij}=p^iL^j+L^ip^j,\quad Y_{20}^{ij}=(\vec{\sigma},\vec{p})(x^ip^j+p^ix^j),\quad Y_{21}^{ij}=(\vec{x},\vec{p})(L^i\sigma^j+\sigma^i L^j),\nonumber\\ &
Y_{22}^{ij}=(\vec{\sigma},\vec{x})(L^iL^j+L^jL^i),\quad Y_{23}^{ij}=(\vec{x}\wedge\vec{\sigma})^iL^j+L^i(\vec{x}\wedge\vec{\sigma})^j,\quad Y_{24}^{ij}=(\vec{x},\vec{p})(\vec{\sigma},\vec{p})x^ix^j,\nonumber\\
&Y_{25}^{ij}=(\vec{x},\vec{p})^2(x^i\sigma^j+\sigma^ix^j),\quad Y_{26}^{ij}=(\vec{x},\vec{p})^2(\vec{\sigma},\vec{x})x^ix^j.
\end{align}
Each of the quantities in \eqref{pseudo} can be multiplied by a scalar $f(r)$ without changing its properties under rotations or reflections. 

It can be shown that we have one linear relation
\begin{equation}\label{peq0}
Y_{20}^{ij}=Y_9^{ij}+Y_{16}^{ij}-Y_{11}^{ij}-Y_{13}^{ij}, 
\end{equation}
and even though all the above pseudo-tensors except $Y_{20}^{ij}$ are linearly independent, higher-order polynomial relations between them exist. 
Indeed, we have the following syzygies: 
\begin{align}
&Y_{21}^{ij}=-\vec{r}^{\:2}Y_{13}^{ij}+Y_{12}^{ij}+Y_{18}^{ij},\nonumber\\
&Y_{22}^{ij}=-\vec{r}^{\:2}Y_{16}^{ij}+2Y_{15}^{ij}-Y_{21}^{ij},\nonumber\\
&Y_{23}^{ij}=-\vec{r}^{\:2}Y_{3}^{ij}+Y_{2}^{ij}-Y_{5}^{ij}+Y_{7}^{ij},\nonumber\\
&Y_{24}^{ij}=-\vec{r}^{\:2}(Y_{13}^{ij}+Y_{11}^{ij}-Y_{9}^{ij})+Y_{18}^{ij}-Y_{10}^{ij}+Y_{15}^{ij}-Y_{21}^{ij},\nonumber\\
&Y_{25}^{ij}=\vec{r}^{\:2}Y_{9}^{ij}-Y_{10}^{ij},\nonumber\\
&Y_{26}^{ij}=\vec{r}^{\:2}Y_{22}^{ij}-Y_{14}^{ij}\label{peq1}. 
\end{align}
We use relations \eqref{peq0} and \eqref{peq1} to remove the left-hand sides
of these equations from the analysis completely.
\subsection{The commutativity condition and determining equations}\label{sec4.1}
Here, we take the linear combinations of all the two-component pseudo-tensors 
and then fully symmetrize them in order to obtain the determining equations from the commutativity condition.

Let us take the linear combination of the independent pseudo-tensors given in \eqref{pseudo} as 
\[X_Y^{ij}=\sum\limits_{k=1}^{19}f_k(r)Y_{k}^{ij}, \]
which can be fully symmetrized as decribed in \cite{DWY}. Similar to the tensors it is sufficient to consider the full symmetric 
form of $X_Y^{12}$. From the requirement $[H,X_Y^{ij}]=0$, we obtain the determining equations.

The determining equations obtained by equating the coefficients of third-order terms to zero in the commutativity equation, are given as follows:
\begin{align}
& f_{17}=0, \quad f_{19}=c_1,\quad f_{10}=f_{12}-f_{18},\quad
f_{9}=-f_{11}, \label{peq4} \medskip \\ & f_{13}=f_{11}-r^2f_{12}+c_2,\quad f_{11}=r^2f_{15}+c_3,\quad
f_{16}=-r^2 f_{15}+\dfrac{c_4}{r},\label{peq5} \medskip \\
&2rf_{14}V_1+\hbar f_{14}'=0, \label{peq6}\\
&\hbar f_{12}+2c_2V_1+\hbar r f_{12}'=0, \label{peq7}\\
&\hbar rf_{14}+2r f_{18}V_1+\hbar(f_{12}'+r^2f_{14}')=0, \label{peq8}\\
&2rf_{12}V_1+\hbar f_{18}'=0, \label{peq9}\\
&2\hbar(c_4-r^3(f_{15}+f_{18}))-4r^2(c_4(c_2+c_3)r-r^3f_{12})V_1=0, \label{peq10}\\
&\hbar(f_{12}+2r^2f_{14}+f_{15}-f_{18})-2(c_3+r^2(-f_{12}+f_{15}+f_{18}))V_1=0, \label{peq11}\\
&-2c_3V_1+f_{12}(\hbar+4r^2V_1)+\hbar(f_{15}-f_{18}+r(f_{12}'+f_{15}'+f_{18}'))=0,
\label{peq12}
\end{align}
where $c_i\,(i=1,2,3,4)$ are integration constants. \\
Introducing the relations given in
\eqref{peq4} and \eqref{peq5} into the determining equations obtained by equating the coefficients of the second-order
terms to zero in the commutativity equation, we get \allowdisplaybreaks
\begin{align}
& f_2'+c_1V_1'=0, \label{peq13}\\
& f_2-c_1V_1+rf_2'=0,\label{peq14}\\
& rf_2+c_1rV_1+f_3'=0,\label{peq15}\\
& \hbar f_7+2(-c_1\hbar+f_3)V_1=0,\label{peq16}\\
& 2\hbar f_5+2\hbar r^2 f_6+\hbar f_7+2c_1\hbar V_1+2r^2f_2V_1-2f_3V_1+\hbar rf_7'=0, \label{peq17}\\
&-2\hbar rf_6-2r(f_2+f_7)V_1-\hbar f_7'+c_1\hbar V_1'=0,\label{peq18}\\
&2 r f_{6} V_1+{\hbar} f_{6}'=0,\label{peq19}\\
& 4\hbar rf_6+2r(f_2+2f_5+f_7)V_1+\hbar(-f_2'+f_7')=0,\label{peq20n}\\
& \hbar(f_2+2f_5+f_7)+(c_1\hbar-2(f_3+r^2f_7))V_1=0,\label{peq20n2}\\
& 2r(-f_2+f_7)V_1+2rf_6(3\hbar-2r^2V_1)+\hbar(f_2'+2f_5'+f_7')=0,\label{peq20n3}\\
& 4\hbar rf_6+2r(f_2+2f_5+f_7)V_1+\hbar(f_7'+c_1V_1')=0,\label{peq20}\\
&2\hbar rf_6-2r f_2V_1+\hbar (f_5'+r^2f_6')=0,\label{peq21}\\
&rf_8+(2c_4+3(c_2+c_3)r+r^3(-3f_{12}+f_{15}+f_{18}))V_1+c_2r^2V_1'=0, \label{peq22}\\
&2\hbar rf_6+2r(f_2+f_7)V_1+\hbar(-2f_2'+f_7')-3c_1\hbar V_1'=0, \label{peq23} \\
&2r(f_2+2f_5-f_7)V_1+4rf_6(-\hbar+r^2V_1)-\hbar(2f_2'+2f_5'+f_7')+c_1\hbar V_1'=0. \label{peq24}
\end{align}
In like manner, introducing the relations given in
\eqref{peq4} and \eqref{peq5} into the determining equations obtained by equating the coefficients of the first-order
terms to zero in the commutativity equation yields \allowdisplaybreaks 
\begin{align}
& 4r(r(2f_5-f_7)V_1+c_1V_0')+2r(-c_5+2c_6r^2+2f_3)V_1'=0, \label{peq25}\\
& 6 c_4 \hbar^3-4 \hbar r^5 f_4+8 \hbar^2 r^5 f_{14} (\hbar+2 r^2 V_1)\nonumber\\ &\quad+
r^2 \bigg(2 V_1 \Big(-3 c_4 \hbar^2+r^3 \pig(-2 f_{1}+\hbar \big(2 f_{8}+\hbar (f_{12}+5
(f_{15}+f_{18})-r f_{15}')\big)\pig)\Big)\nonumber\\ &\quad
+\hbar r (12 \hbar^2 r f_{12}'+20 \hbar^2 r^3 f_{14}'+14 \hbar^2 r f_{15}'+4 c_4 V_0'+2
c_2 r V_0'+2 c_3 r V_0'+2 c_4 \hbar V_1'+3 c_2 \hbar r V_1'\nonumber\\ &\quad-c_3 \hbar
r V_1'-2 \hbar r^3 f_{15} V_1'+ 
2 \hbar^2 r^2 f_{12}''+2 \hbar^2 r^4 f_{14}''+2 \hbar^2 r^2 f_{15}'')\bigg) = 0, \label{peq26}\\
&4 c_4 \hbar^2+4 r^3 f_{1}-8 \hbar r^5 f_{14} (\hbar-r^2 V_1)+
r^2 \Big(2 \big(\hbar (-4 c_4+c_2 r-c_3 r)+r^3 (5 \hbar f_{12}+\hbar f_{18}+2 f_{8})\big)
V_1\nonumber\\ &\quad+2 \hbar r f_{15} (-2 \hbar+5 r^2 V_1)+ 
r^2 \pig(4 c_2 V_0'+\hbar \big(4 c_2 V_1' +\hbar (6 f_{12}'+6 f_{15}'+6
f_{18}'\nonumber\\ &\quad+r (f_{12}''+f_{15}''+f_{18}''))\big)\pig)\Big) = 0, \label{peq27}\\
&2(-2c_4\hbar+r^3(13\hbar f_{12}+7\hbar f_{15}-7\hbar f_{18}+2f_8))V_1-40\hbar r^3f_{14}(\hbar-r^2V_1)+r\big(-4rf_1'-6\hbar^2 rf_{12}' \nonumber\\
&\quad+10\hbar^2 rf_{15}'+18\hbar^2 rf_{18}'+4c_2rV_0'+8c_4\hbar V_1'+4\hbar r(c_2+c_3)V_1'+\hbar^2 r^2(-f_{12}''+f_{15}''+3f_{18}'')  \big)=0, \label{peqq28}\\
&-12 c_4 \hbar^3+8 \hbar^2 r^5 f_{14} (\hbar-15 r^2 V_1)+16 r^5 f_4 (\hbar-r^2
V_1)+r^2 (V_1 (8 c_4 \hbar^2+2 r^3 (-4 f_1+\hbar (2 f_8 \nonumber\\&\quad+\hbar
(-27 f_{12}-13 f_{15}+17 f_{18}+2 r f_{15}'))))+\hbar r (4 r f_1'+30
\hbar^2 r f_{12}'-80 \hbar^2 r^3 f_{14}'-6 \hbar^2 r f_{15}'\nonumber \\&\quad-42 \hbar^2 r f_{18}'-8 c_4
V_0'+4 c_3 r V_0'+8 r^3 f_{15} V_0'-4 c_4 \hbar V_1'-6 c_2 \hbar r V_1'-14
c_3 \hbar r V_1'+4 \hbar r^3 f_{15} V_1'\nonumber\\&\quad+5 \hbar^2 r^2 f_{12}''-8 \hbar^2 r^4 f_{14}''+15
\hbar^2 r^2 f_{15}''-7 \hbar^2 r^2 f_{18}''+8 c_4 \hbar r V_1''+2 \hbar^2 r^3 f_{15}^{(3)}))=0.\label{peqq29}
\end{align}
Finally, introducing the relations given in
\eqref{peq4} and \eqref{peq5} into the determining equations obtained by equating the coefficients of the zeroth-order
terms to zero in the commutativity equation, we obtain \allowdisplaybreaks
\begin{align}
& 8r^4f_4V_1-\hbar(4c_4V_0'+r^3(8f_4'-4f_{15}V_0'+rf_4'')-2c_4rV_0'') =0,\label{peq28}\\
& 2r^2f_1V_1+4r^2f_4(-\hbar+r^2V_1)-\hbar(4rf_1'+(-2c_4-c_2r+c_3r+2r^3f_{15})V_0'+r^2f_1'') =0,\label{peq29}\\
& 2\hbar(-3\hbar+4r^2V_1)f_5'+56\hbar^2r^2f_6'+8\hbar r^4V_1f_6'-6\hbar^2f_7'+8\hbar r^2V_1f_7'+4r^2V_0'(f_5+f_7)  \nonumber\\ &\quad +
4r^3f_6(12\hbar V_1+rV_0')+6\hbar^2rf_5''+16\hbar^2r^3f_6''+6\hbar^2rf_7''+\hbar^2r^2(f_5^{(3)}+r^2f_6^{(3)}+f_7^{(3)}) =0,\\
&2 c_5 \hbar^2+24 \hbar r^4 f_6 (\hbar-r^2 V_1)+ r \big(-2 \hbar^2 f_3'-2 \hbar r V_1 \Big(-c_5+4 c_6 r^2+r \pig(f_3'+r \big(2 f_5+f_7 \nonumber\\ &\quad  +2
r (f_5'+r^2 f_6'+f_7')\big)\pig)\Big)+r (12 \hbar^2 r f_5'+4 \hbar^2 r^3 f_6'+8 \hbar^2 r f_7'-2 c_5 r V_0'+4
c_6 r^3 V_0'+4 r f_3 V_0'\nonumber\\ &\quad +2 \hbar^2 f_3''+2 \hbar^2 r^2 f_5''+\hbar^2 r^2 f_7''+\hbar^2
r f_3^{(3)})\big) =0.\label{peq31}
\end{align}
Similar to the situation with tensors, we classify the above determining equations into four types:
\begin{enumerate}
	\item  Those that are independent of the potentials $V_1$ and $V_0$, namely \eqref{peq4} and \eqref{peq5}. These relations are substituted into the other determining	equations. This eliminates some of the equations involving the potential and we have presented only the remaining ones.
	\item  Those involving only $V_1$ but not its derivative \eqref{peq6}-\eqref{peq12}, \eqref{peq14}-\eqref{peq17}, \eqref{peq19}-\eqref{peq20n3}, \eqref{peq21}.
	\item Those involving $V_1$ and $V'_1$ which are \eqref{peq13}, \eqref{peq18}, \eqref{peq20}, \eqref{peq22}-\eqref{peq24}.
	\item Those involving  $V_0$, $V_1$ and their derivatives \eqref{peq25}-\eqref{peq31}.
\end{enumerate}

\subsection{Solutions of the determining equations} 
The aim of this subsection is to present the solutions of the determining equations given in section \ref{sec4.1} and then reveal the pseudo-tensor integrals of motion. Similar to the tensors, we shall analyze the determining equations by following the above four classes and again the determining equations of first type had already been substituted to the others to eliminate or simplify them. Hence, we start with the determining equations of second type. But first, one can see that integrating \eqref{peq13} allows us to consider it as a determining equation of second type:
\begin{equation}\label{peqn1}
	f_2+c_1V_1+c_5=0,
\end{equation}
where $c_1$ and $c_5$ are real constants. This equation will simplify the analysis. 

Adding \eqref{peqn1} to \eqref{peq13} yields
\begin{equation*} 
rf_2'+2f_2+c_5=0,
\end{equation*}
which can be integrated to give 
\begin{equation}\label{peq39}
f_2=-\dfrac{c_5}{2}+\dfrac{c_6}{r^2},
\end{equation}
where $c_6$ is an integration constant. Introducing \eqref{peq39} into \eqref{peq14}, we obtain
\begin{equation}\label{peq40}
c_1 V_1+\dfrac{c_6}{r^2}+\dfrac{c_5}{2}=0.
\end{equation}
To continue analyzing, it is better to consider the following cases depending upon the constant $c_1$:\\
\textbf{Case 1.} $c_1\neq0$.\\
In this case, we immediately get a potential as
\begin{equation}\label{peq41}
V_1=-\dfrac{c_6}{c_1r^2}-\dfrac{c_5}{2c_1}.
\end{equation}
Introducing this equation into \eqref{peq15} and integrating, we get
\begin{equation}\label{peq42}
f_3=\dfrac{c_5 r^2}{2}+c_7,
\end{equation}
where $c_7$ is an integration constant. Considering \eqref{peq41} and \eqref{peq42} in equation \eqref{peq16}, we obtain 
\begin{equation}\label{peq43}
f_7=\dfrac{(2c_7-2c_1\hbar+c_5r^2)(2c_6+c_5r^2)}{c_1\hbar r^2}.
\end{equation}
On the other hand, if we introduce \eqref{peqn1}, \eqref{peq41}-\eqref{peq43} into the equation \eqref{peq17}, we find
\begin{equation}\label{peq44}
f_6=\dfrac{4{c_6}^2-4c_5c_6r^2+c_5r^2(-4c_7+4c_1\hbar-5c_5r^2)-4c_1\hbar r^2f_5}{4c_1\hbar r^4},
\end{equation}
which can be considered in \eqref{peq21} to deduce 
\begin{equation}\label{peq45}
c_5=0.
\end{equation}
If we introduce \eqref{peq39}, \eqref{peq41}, \eqref{peq43}, \eqref{peq44} and \eqref{peq45} into \eqref{peq20n}, we obtain
\begin{equation}
\dfrac{(c_6+c_1\hbar)(-2c_6c_7+3c_1c_6\hbar-2c_1\hbar r^2f_5)}{{c_1}^2\hbar r^3}=0.
\end{equation}
Obviously, we encounter two different cases here. However, the case $c_6=-c_1\hbar$ gives the potential $V_1=\frac{\hbar}{r^2}$ which can be induced by a gauge transformation 
and had been considered thoroughly in \cite{wy3}. So, we take
\begin{equation}
	f_5=\dfrac{c_6(-2c_7+3c_1\hbar)}{2c_1\hbar r^2},
\end{equation} 
which together with \eqref{peq44} can be considered in \eqref{peq19} to give
\begin{equation}
\dfrac{c_6(c_6+2c_1\hbar)(-2c_6-2c_7+3c_1\hbar)}{{c_1}^2\hbar r^5} =0.
\end{equation}
Then we have three subcases. The first one is the case $c_6=0$. However, it leads to $V_1 = 0$. Therefore, we analyze the following two subcases.
\begin{itemize}
	\item[]\textbf{Subcase 1.} $c_6=-2c_1\hbar$. \\
	From \eqref{peq41} and \eqref{peq45}, we immediately obtain 
\begin{equation}\label{peq48}
V_1=\dfrac{2\hbar}{r^2}.
\end{equation}
Then, from \eqref{peq20n2} it is easy to show that
\begin{equation}
	c_7=\dfrac{3c_1\hbar}{2}.
\end{equation}

By considering the above spin-orbital potential, let us solve all the determining equations of second type. Integrating \eqref{peq6} and \eqref{peq7}, we find
\begin{equation}\label{peq49}
f_{14}=\dfrac{k_1}{r^4},\quad f_{12}=\dfrac{4c_2+r k_2}{r^2},
\end{equation}
where $k_1$ and $k_2$ are integration constants.
Then, from \eqref{peq8} we get
\begin{equation}\label{peq50}
f_{18}=\dfrac{8c_2+3k_1+k_2 r}{4r^2}.
\end{equation}
Introducing \eqref{peq49} and \eqref{peq50} into \eqref{peq9} yields
\begin{equation}\label{peq51}
k_1=8c_2,\quad k_2=0.
\end{equation}
By substituting \eqref{peq49}, \eqref{peq50} and \eqref{peq51} into the equations \eqref{peq10}, \eqref{peq11} and \eqref{peq12}, one can conclude that
\begin{equation}\label{peq52}
f_{15}=-\dfrac{4(c_2+c_3)}{3r^2},\quad c_3=2c_2,\quad c_4=0.
\end{equation}
Hence, all the determining equations of second type are satisfied. Thus, we continue the analysis with the determining equations of third type. 
	
Substituting \eqref{peq48}, \eqref{peq49}, \eqref{peq50} and \eqref{peq52} into \eqref{peq22} gives us
\begin{equation}
f_8 = \dfrac{2c_2\hbar}{r^2},
\end{equation}
which guarantees that the determining equations of third type are also satisfied. Introducing the above relations into the remaining determining equations yields
 \begin{equation}
 V_0=\dfrac{3\hbar^2 }{r^2},\quad f_1=-\dfrac{9c_2\hbar^2}{r^2},\quad f_4=\dfrac{24c_2\hbar^2}{r^4},
 \end{equation}
 where a redundant additive constant in $V_0$ is omitted. 

 Hence, all determining equations are satisfied. For this case, we have two arbitrary constants $c_1$ and $c_2$. The two integrals of motion are:
 	 	\begin{align} 
 	\mathcal{Y}_1^{\,ij}=&p^iL^j+L^ip^j-\dfrac{2\hbar}{r^2}\big(\mathrm{i}\hbar+(\vec{x},\vec{p})\big)(x^i\sigma^j+\sigma^ix^j)+\dfrac{3\hbar}{2}(p^i\sigma^j+\sigma^ip^j)\nonumber\\
 	&+\dfrac{4\hbar}{r^4}\big(3\mathrm{i}\hbar + 2(\vec{x},\vec{p})\big)(\vec{\sigma},\vec{x})x^ix^j-\dfrac{2\hbar}{r^2}(\vec{\sigma},\vec{x})(x^ip^j+p^ix^j),\label{eqY1}\\
 	\mathcal{Y}_2^{\,ij}=&-2\big( \dfrac{2\hbar^2}{r^2}-\dfrac{2\mathrm{i}\hbar}{r^2}(\vec{x},\vec{p})-\vec{p}^{\:2}+\dfrac{2}{r^2}\vec{L}^{\:2}\big)(x^i\sigma^j+\sigma^ix^j)-\big(4\mathrm{i}\hbar+2(\vec{x},\vec{p})\big)(p^i\sigma^j+\sigma^ip^j)\nonumber \\
 	&-\dfrac{8}{r^2}\big(\dfrac{\mathrm{i}\hbar}{r^2}(\vec{x},\vec{p})(\vec{\sigma},\vec{x})-\dfrac{2\hbar^2}{r^2}(\vec{\sigma},\vec{x})+\mathrm{i}\hbar(\vec{\sigma},\vec{p})-\dfrac{2}{r^2}(\vec{\sigma},\vec{x})\vec{L}^{\:2}\big)x^ix^j+8(\vec{\sigma},\vec{x})p^ip^j\nonumber\\
 	&+\dfrac{4}{r^2}(\vec{x},\vec{p})\big((\vec{x}\wedge\vec{\sigma})^iL^j+L^i(\vec{x}\wedge\vec{\sigma})^j \big)-5\big((\vec{p}\wedge\vec{\sigma})^iL^j+L^i(\vec{p}\wedge\vec{\sigma})^j\big)\nonumber\\
 	&-\dfrac{4}{r^2}(\vec{x},\vec{p})(\vec{\sigma},\vec{x})(x^ip^j+p^ix^j)+\dfrac{2}{r^2}\big(\hbar+4(\vec{\sigma},\vec{L})\big)(x^iL^j+L^ix^j).\label{eqY2}
 	\end{align}
 \begin{rem}
Note that only the off-diagonal elements of the above pseudo-tensors commute with the Hamiltonian. Observe that they are not traceless and none of the diagonal elements are integrals of motion. However, we know that traces of such integrals of motion appear separately as (pseudo)scalars. Indeed, if the diagonal elements are redefined as $\mathcal{Y}^{ii}-\frac{1}{3}\,tr(\mathcal{Y}^{ij})$, then these new elements commute with the Hamiltonian as well. For example, the trace of $\mathcal{Y}_1^{\,ij}$ is the pseudo-scalar $3(\vec{\sigma},\vec{p})$ and so $\mathcal{Y}_1^{\,ii}-(\vec{\sigma},\vec{p})$ commutes with the Hamiltonian. Note that in the rest of the paper only one pseudo-tensor integral of motion $\mathcal{Y}_6^{\,ij}$ will be traceless. 

 \end{rem}
	\item[]\textbf{Subcase 2.} $-2c_6-2c_7+3c_1\hbar=0$.\\
	By using the relation $c_7=-c_6+\dfrac{3c_1\hbar}{2}$ in \eqref{peq20n2}, one can directly obtain
	\[ \dfrac{{c_6}^2}{{c_1}^2}-\hbar^2=0, \]
	which gives either $c_6=-c_1\hbar$ or $c_6=c_1\hbar$. However, the first one is  excluded due to the same reason mentioned before. 
	Then, we continue to analyze by taking $c_6 = c_1\hbar$, which in the vicinity of \eqref{peq41} gives 
	\begin{equation}\label{peq55}
	V_1=-\dfrac{\hbar}{r^2}.
	\end{equation}
	By following very similar steps to Subcase 1, one can obtain the relations given as: 
	\begin{equation}\label{peq56}
	f_{14}=0,\quad f_{12}=-\dfrac{2c_2}{r^2},\quad f_{18}=\dfrac{2 c_2}{r^2},\quad c_3=-2c_2,\quad c_4=0,\quad f_{15}=0.
	\end{equation}
   With these relations all the determining equations of second type are satisfied.
	On the other hand, introducing the relations \eqref{peq55} and \eqref{peq56} into \eqref{peq22} we get 
	\begin{equation*}
	f_8=\dfrac{3c_2\hbar}{r^2}.
	\end{equation*}
        Then, the determining equations of third type are also satisfied. 
        Finally, considering the above relations in the remaining determining equations give us 
	\begin{equation*}
	V_0=\alpha,\quad f_1=-\dfrac{3c_2\hbar^2}{2r^2},\quad f_4=0,
	\end{equation*}
where $\alpha$ is a real constant.
	Therefore, all the determining equations are satisfied for the spin-orbital potential $V_1=-\frac{\hbar}{r^2}$ and the scalar one $V_0=\alpha$. Having left with two arbitrary constants $c_1$ and $c_2$, there exist two integrals of motion which read:
	 	 	\begin{align}
	\mathcal{Y}_3^{\,ij}=&p^iL^j+L^ip^j+\dfrac{\hbar}{r^2}\big(\mathrm{i}\hbar+(\vec{x},\vec{p})\big)(x^i\sigma^j+\sigma^ix^j)+\dfrac{\hbar}{2}(p^i\sigma^j+\sigma^ip^j)\nonumber\\
	& -\dfrac{\hbar}{r^2}(\vec{\sigma},\vec{x})(x^ip^j+p^ix^j) +\dfrac{2\hbar}{r^2}(\vec{\sigma},\vec{p})x^ix^j,\label{eqY3}\medskip\\
	\mathcal{Y}_4^{\,ij}=&(\dfrac{3\hbar^2}{2r^2}-\dfrac{2\mathrm{i}\hbar}{r^2}(\vec{x},\vec{p})+2\vec{p}^{\:2}-\dfrac{4}{r^2}\vec{L}^{\:2})(x^i\sigma^j+\sigma^ix^j)+2(\mathrm{i}\hbar-(\vec{x},\vec{p}))(p^i\sigma^j+\sigma^ip^j)\nonumber\\ &+\dfrac{2\mathrm{i}\hbar}{r^2}\big( 2(\vec{\sigma},\vec{p})x^ix^j-(\vec{\sigma},\vec{x})(x^ip^j+p^ix^j) \big)-\dfrac{2}{r^2}(\vec{x},\vec{p})\big( (\vec{x}\wedge\vec{\sigma})^iL^j+L^i (\vec{x}\wedge\vec{\sigma})^j\big)\nonumber\\
	&+\big((\vec{p}\wedge\vec{\sigma})^iL^j+L^i (\vec{p}\wedge\vec{\sigma})^j\big)+\dfrac{1}{r^2}(2(\vec{\sigma},\vec{L})+3\hbar)(x^iL^j+L^ix^j)\,.\label{eqY4}
	\end{align}
\end{itemize}
\textbf{Case 2.} $c_1=0$, $V_1$ unspecified.\\
In this case, \eqref{peq14} together with \eqref{peq39} yield 
\begin{equation}\label{peq58}
c_5=0, \quad c_6=0.
\end{equation}
 Introducing these into \eqref{peq15} and using \eqref{peq39}  give
\begin{equation}
f_3=c_8,\quad c_8\in\mathbb{R}.
\end{equation} 
By considering these relations in \eqref{peq21} and then integrating, we find
\begin{equation}\label{peq60}
f_5=-r^2f_6+c_9,
\end{equation}
where $c_9$ is an integration constant. 
Introducing this equation into the sum of \eqref{peq16} and \eqref{peq17}, we obtain
\begin{equation}\label{peq61}
f_7=-c_9+\dfrac{c_{10}}{r^2},\quad c_{10}\in\mathbb{R}, 
\end{equation}
which can be substituted back into \eqref{peq16} to obtain
\begin{equation}\label{peq62}
2c_8V_1+\dfrac{\hbar c_{10}}{r^2}-\hbar c_9=0.
\end{equation}
If $c_8\neq 0$, we have
\begin{equation}\label{peq63}
V_1=\dfrac{\hbar(c_9 r^2-c_{10})}{2c_8r^2}.
\end{equation} 
Introducing \eqref{peq58}, \eqref{peq60}, \eqref{peq61} and \eqref{peq63} into \eqref{peq20n} and \eqref{peq20n3} and after making some computations we obtain 
\begin{equation*}
c_9=0,\quad f_6=0,\quad c_{10}(2c_8+c_{10})=0\,,
\end{equation*}
which implies the existence of two subcases, namely $c_{10}=0$ or $c_{10}=-2c_8$ to consider. However, the former yields $V_1=0$ and the latter gives the gauge induced potential $V_1=\frac{\hbar}{r^2}$.

Thus, we take $c_8=0$ which immediately yields $c_9=0$ and $c_{10}=0$. Then, by \eqref{peq20n} we find $f_6=0$. We continue the analysis by considering the determining equation \eqref{peq7} which now reads 
\begin{equation}\label{peqlast}
	\hbar f_{12}+2c_2V_1+\hbar r f_{12}'=0.
\end{equation}
Let us examine the following subcases based upon the constant $c_2$. 
\begin{itemize}
	\item[]\textbf{Subcase 1.} $c_2=0$. \\
This case directly yields
\begin{equation}\label{peqy1}
	f_{12}=\dfrac{c_{11}}{r},
\end{equation}
where $c_{11}$ is a real constant.
One can show after a routine computation that the case $c_{11}=0$ gives either $V_1=0$ or $V_1=\frac{\hbar}{r^2}$. Hence, we assume $c_{11}\neq 0$. Then, from \eqref{peq9} we obtain
\begin{equation}\label{peqy2}
	V_1=-\dfrac{\hbar f_{18}'}{2c_{11}}.
\end{equation}
Introducing these into \eqref{peq12}, \eqref{peq10} and \eqref{peq8}, respectively, we find
\begin{align}
	&f_{18}=\dfrac{c_{12}-rc_{11}f_{15}}{c_3-rc_{11}},\quad
	f_{15}=-\dfrac{c_3 c_{11}(c_4+rc_{13})-r(c_4({c_{11}^2}+c_{12})+r{c_{11}^2}c_{13})}{r^2 c_{11}(c_4+r(c_3-rc_{11}))},\label{peqy3}\\
	&f_{14}=-\dfrac{(c_3c_{11}-r(c_{11}^2+c_{12})-c_{11}c_{13})(2c_4c_{11}+r(c_3c_{11}-rc_{11}^2+rc_{12}+c_{11}c_{13}))}{2r^2c_{11}(c_4+r(c_3-rc_{11}))^2}+\dfrac{c_{14}}{r},\label{peqy4}
\end{align}
where $c_{i}$ ($i=12,13,14$) are integration constants. If the relations \eqref{peqy1}-\eqref{peqy4} are introduced into the remaining determining equations of second type, the following relations are obtained:
\begin{equation}
	c_{14}=0,\quad (3c_{11}^2+c_{12})(c_{11}^2+c_{12})(c_{11}^2-c_{12})=0.
\end{equation}
Therefore, we have the following three possibilities:\medskip\\
	\textbf{I.} $c_{12}=-3c_{11}^2$.\\
	This immediately gives us the potential 
	\[V_1=\dfrac{3\hbar}{2r^2}.\]
	Then, we obtain
	\begin{equation}
		c_{13}=0,\quad c_3=0,\quad c_4=0.
	\end{equation}
	Hence, the determining equations of second type are satisfied. By introducing these relations into the remaining equations, we get
	\begin{equation*}
	f_8=0,\quad  f_1=-\dfrac{6\hbar^2c_{11}}{r},\quad f_4=\dfrac{15\hbar^2 c_{11}}{r^3}.
	\end{equation*}
	Now, all the determining equations are satisfied for any $V_0=V_0(r)$. Having just one arbitrary constant $c_{11}$ we have the following integral of motion:
	\begin{align}
	\mathcal{Y}_5^{\,ij}=&-r\big((\vec{p}\wedge\vec{\sigma})^iL^j+L^i (\vec{p}\wedge\vec{\sigma})^j\big)+\dfrac{(\vec{x},\vec{p})}{r}\big((\vec{x}\wedge\vec{\sigma})^iL^j+L^i (\vec{x}\wedge\vec{\sigma})^j\big)\nonumber\\
	&+\dfrac{3(\vec{\sigma},\vec{L})}{r}(x^iL^j+L^ix^j)-\mathrm{i}\hbar r(p^i\sigma^j+\sigma^ip^j)+\dfrac{\mathrm{i}\hbar}{r}(\vec{\sigma},\vec{x})(x^ip^j+p^ix^j)\label{eqY5}\\
	&+\dfrac{2}{r}\pig(\dfrac{(\vec{\sigma},\vec{x})}{r^2}\big(4\vec{L}^{\:2}+\hbar^2\big)-\mathrm{i}\hbar(\vec{\sigma},\vec{p}) \pig)x^ix^j+\dfrac{1}{r}\pig(-2\vec{L}^{\:2}+\mathrm{i}\hbar(\vec{x},\vec{p})-\dfrac{3\hbar^2}{2}\pig)(x^i\sigma^j+\sigma^ix^j)\nonumber.
	\end{align}
	\medskip\\
	\textbf{II.} $c_{12}=-c_{11}^2$.\\
	In this case, we directly find
	\[ V_1=\dfrac{\hbar}{2r^2}. \]
	Similar to the above case, introducing these into the remaining determining equations yields
	\begin{align*}
	c_{13}=0,\quad c_3=0,\quad c_4=0,\quad f_8=\dfrac{\hbar c_{11}}{r},\quad  f_1=0,\quad f_4=0.
	\end{align*}
	So, all the determining equations are satisfied for arbitrary $V_0=V_0(r)$. We have just one arbitrary constant $c_{11}$ which yields the following integral of motion:
	\begin{align}
	\mathcal{Y}_6^{\,ij}=&-r\big((\vec{p}\wedge\vec{\sigma})^iL^j+L^i (\vec{p}\wedge\vec{\sigma})^j\big)+\dfrac{(\vec{x},\vec{p})}{r}\big((\vec{x}\wedge\vec{\sigma})^iL^j+L^i (\vec{x}\wedge\vec{\sigma})^j\big)\nonumber\\
	&+\dfrac{1}{r}\big((\vec{\sigma},\vec{L})+\hbar\big)(x^iL^j+L^ix^j)-\mathrm{i}\hbar r(p^i\sigma^j+\sigma^ip^j)+\dfrac{\mathrm{i}\hbar}{r}(\vec{\sigma},\vec{x})(x^ip^j+p^ix^j)\nonumber\\
	&-\dfrac{2\mathrm{i}\hbar}{r}(\vec{\sigma},\vec{p})x^ix^j+\dfrac{\hbar}{r}\pig(\mathrm{i}(\vec{x},\vec{p})-\dfrac{\hbar}{2}\pig)(x^i\sigma^j+\sigma^ix^j).\label{eqY6}
	\end{align}
However, the above pseudo-tensor integral of motion can be written as
	\[\mathcal{Y}_6^{\,ij}=\{X_V^i,\mathcal{J}_j \}+\delta_{ij}\dfrac{\hbar^2 X_P}{2}, \]
where
\[ X_P=\dfrac{(\vec{\sigma},\vec{x})}{r} \]
is a first-order pseudo-scalar integral of motion and
\[ \vec{X}_V=\{\vec{\mathcal{J}},X_P\} \]
is a first-order vector integral of motion for the case $V_1=\dfrac{\hbar}{2r^2}$ and $V_0=V_0(r)$, (For details see; \cite{wy3,DWY}). Hence, we say that $\mathcal{Y}_6^{\,ij}$ is an obvious integral of motion.
\medskip\\
	\textbf{III.} $c_{12}=c_{11}^2$.\\
	This case yields the potential 
	\[V_1=-\dfrac{\hbar}{2r^2}. \]
	By following similar steps to the above cases, we find the following relations:
	\begin{align*}
	c_{13}=0,\quad c_3=0,\quad c_4=0,\quad f_8=-\dfrac{2\hbar c_{11}}{r},\quad f_1=0,\quad f_4=0.
	\end{align*}
	Again, all the determining equations are satisfied for arbitrary scalar potential $V_0=V_0(r)$ and having just one arbitrary constant $c_{11}$, we obtain the following integral of motion:
	\begin{align}
	\mathcal{Y}_7^{\,ij}=&-r\big((\vec{p}\wedge\vec{\sigma})^iL^j+L^i (\vec{p}\wedge\vec{\sigma})^j\big)+\dfrac{(\vec{x},\vec{p})}{r}\big((\vec{x}\wedge\vec{\sigma})^iL^j+L^i (\vec{x}\wedge\vec{\sigma})^j\big)\nonumber\\
	&-\dfrac{1}{r}\big((\vec{\sigma},\vec{L})+2\hbar\big)(x^iL^j+L^ix^j)+\mathrm{i}\hbar \big(\dfrac{(\vec{\sigma},\vec{x})}{r}-r\big)(p^i\sigma^j+\sigma^ip^j)\nonumber\\
	&-\dfrac{2\mathrm{i}\hbar}{r}(\vec{\sigma},\vec{p})x^ix^j+\dfrac{1}{r}\pig(2\vec{L}^{\:2}+\mathrm{i}\hbar(\vec{x},\vec{p})-\dfrac{3\hbar^2}{2}\pig)(x^i\sigma^j+\sigma^ix^j).\label{eqY7}
	\end{align}
	\item[]\textbf{Subcase 2.} $c_2\neq0$.\\
	With this choice, the equation \eqref{peqlast} directly gives
	\[ V_1=-\dfrac{\hbar(f_{12}+rf_{12}')}{2c_2},\]
and upon introducing this into \eqref{peq9}, we get
\begin{equation}\label{peqlast1}
	f_{18}=\dfrac{r^2 f_{12}^2}{2c_2}+c_{15},
\end{equation}
where $c_{15}$ is an integration constant. Using \eqref{peq6}, \eqref{peq8} and \eqref{peq12}, we obtain
\begin{align}
	&f_{15}=\dfrac{2c_2(rc_{15}+c_{16})-2(c_2+c_3)rf_{12}+r^3f_{12}^2}{2c_2r},\label{peqlast2}\\
	&f_{14}=\dfrac{2c_2rc_{15}f_{12}+r^3f_{12}^3-2c_2(c_2-r^2c_{15})f_{12}'+r^4f_{12}^2f_{12}'}{2c_2r(c_2+r^2f_{12}+r^3f_{12}')}\label{peqlast3},
\end{align}
where $c_{16}$ is a real constant. Note that the denominator of \eqref{peqlast3} is assumed to be nonvanishing. Otherwise it leads to $f_{12}=\dfrac{c_2}{r^2}+\dfrac{a_1}{r}$ for a constant $a_1$ and gives the potential $V_1=\dfrac{\hbar}{2r^2}$. In order to satisfy the other determining equations we need to set $c_2 = 0$ which of course contradicts with the beginning assumption that $c_2\neq 0$. 

By substituting \eqref{peqlast1} and \eqref{peqlast2} in the \eqref{peq10}, we find 
\begin{equation}
	f_{12}=\dfrac{c_4+(c_2+c_3)r+\epsilon\sqrt{c_4^2+2c_3c_4r+r^2(c_2^2+c_3^2+2c_2(c_3-r(rc_{15}+c_{16}))+c_{17})}}{r^3},
\end{equation}
where $c_{17}$ is a real constant and $\epsilon^2=1$. After introducing these relations into the remaining determining equations of second type, we obtain the following relations:
\begin{equation}
c_4=0,\quad c_{16}=0,\quad	c_{15}(c_2^2-c_3^2)(2c_2c_3+c_3^2+c_{17})=0.
\end{equation}
Here, we have three cases to consider.  

The case $c_{15}=0$ yields $c_{17}=-c_2^2-2c_2c_3-c_3^2$ by the help of \eqref{peq11}. This implies that $f_{12}=\dfrac{c_2+c_3}{r^2}$ and $f_{14}=\dfrac{-c_2^2+c_3^2}{2c_2r^4}$. If we introduce these relations into the other determining equations, we find $c_3=\pm c_2$. Then, it is easy to see that the option $c_3=c_2$ gives $V_1=\frac{\hbar}{r^2}$ while the other one $c_3=-c_2$ causes the potential $V_1$ to be vanished.

The case $c_3^2-c_2^2=0$ gives $c_{15}=0$. Then, we are back in the above case.

Finally, for the possibility $2c_2c_3+c_3^2+c_{17}=0$ we easily find that either $c_3=0$ or $c_3=\pm 2c_2$. If $c_3=0$, then we get $c_{15}=0$ which yields either $f_{12}=0$ or $f_{12}=\dfrac{2c_2}{r^2}$ both of which give the excluded potentials. If $c_3=2c_2$, then we obtain $c_{15}=0$. This directly yields either $f_{12}=\dfrac{2c_2}{r^2}$ or $f_{12}=\dfrac{4c_2}{r^2}$. However, we do not need to continue the analysis here since the first case yields a known potential $V_1=\dfrac{\hbar}{r^2}$ and the second one gives $V_1=\dfrac{2\hbar}{r^2}$ which has already been investigated in Subcase 1 of Case 1. Similarly, the case $c_3=-2c_2$ yields either $V_1=0$ or $V_1=-\dfrac{\hbar}{r^2}$ which has been analyzed in Subcase 2 of Case 1.

The above facts state that no new pseudo-tensor integrals of motion are found in this case.
\end{itemize}

\section{Conclusions}
A classification of superintegrable systems preserving rotational invariance, involving spin interaction and admitting second-order integrals of motion was pursued in \cite{DWY}. However, the results presented in \cite{DWY} were restricted to scalar, pseudo-scalar, vector and axial vector integrals of motion. This paper serves as a continuation of \cite{DWY} by considering integrals that are two index tensors and pseudo-tensors. Hence, the classification of such superintegrable systems is completed.

After a complete analysis, we found 6 different tensor integrals which can be seen in \eqref{intmot}, \eqref{eq45}--\eqref{eq49}. However, the first one exists for all $V_0(r)$ and $V_1(r)$ whereas the others correspond to the gauge induced potential $V_1=\frac{\hbar}{r^2}$. Among the tensor integrals for the gauge induced potential, \eqref{eq45}, \eqref{eq47} and \eqref{eq49} exist for any scalar potential $V_0(r)$. However, the integrals \eqref{eq46} and \eqref{eq48} exist only for the scalar potential $V_0(r)=\frac{\hbar^2}{r^2}+\alpha r^2$. Such a system can be viewed as a deformation of the Harmonic oscillator ($V_0(r)=\alpha r^2$). This naturally should bring to mind a well-known integral of motion, the quadrupole tensor \cite{Jauch} (also known as the Fradkin tensor \cite{Fradkin}) for the harmonic oscillator. Notice that in the limit $\hbar \rightarrow 0$ the tensor integral \eqref{eq46} gives the Fradkin tensor for the spinless case whereas \eqref{eq48} vanishes. We conclude that no nontrivial second order tensor integrals of motion exist for the system \eqref{intro1}.  

On the other hand, $12$ different pseudo-tensor integrals of motion have been obtained, seven of which are nontrivial (see \eqref{eqY1}, \eqref{eqY2}, \eqref{eqY3}, \eqref{eqY4}, \eqref{eqY5}--\eqref{eqY7}). However, it has been demonstrated that the integral given in \eqref{eqY6} can be obtained from a first-order pseudo-scalar integral and a first-order vector integral, so it is an obvious integral of motion. 

The other five integrals given in \eqref{eqA6}--\eqref{eqA10} correspond to the gauge induced potential. Among these integrals, \eqref{eqA6}--\eqref{eqA9} do exist for $V_0=\frac{\hbar^2}{r^2}$. There are no pseudo-tensor integrals of motion for arbitrary scalar potential $V_0(r)$. In the case of $V_0=\frac{\hbar^2}{r^2}-\frac{\alpha}{r}$, we have the pseudo-tensor integral \eqref{eqA10}. This superintegrable system represents a deformation of the Kepler–Coulomb one $V_0=\frac{\alpha}{r}$. It is easy to see that \eqref{eqA10} reduces to zero for the spinless case.  

All the results presented in \cite{DWY} and obtained in this paper are summed up in Table \ref{table1} which represents the complete list of superintegrable potentials and their integrals of motion for the current Hamiltonian system under investigation. Column 2 gives the spin-orbital potentials arising from the analysis. In column 3, the scalar potentials are given where $V_0(r)$ means an arbitrary potential. The results obtained in \cite{DWY} are listed in columns 4, 5 and 6. However, we leave blank some entries corresponding to the gauge induced potential in these columns since this potential was excluded from the analysis in \cite{DWY}. As expected there may be some (pseudo)scalar or (axial)vector integrals of motion for these cases but it is not necessary to present them here. Finally, integrals of motion corresponding to superintegrable potentials obtained in this paper are presented in columns 7 and 8. The real constants $\alpha$ and $\beta$ are arbitrary and $\epsilon=\pm 1$. One can conclude from the table that we have four new nontrivial superintegrable potentials given in No. 3, 4, 5 and 6. The remaining one given in No. 2 has been previously found in \cite{DWY} for pseudo-scalars and vectors. We have shown that there is also an obvious pseudo-tensor integral of motion in addition to those for this case. 

A study of the algebras of the integrals of motion presented in this paper is in progress and will be presented in a future article. 

\begin{landscape}
\begin{table}
	\caption{Complete list of superintegrable potentials and their integrals of motion}
	\label{table1}
	\begin{tabular}{ l l l l l l l l }
		\hline
		\noalign{\medskip}
		No & $V_1$ & $V_0$ & Pseudo-scalars & Vectors & Axial Vectors & Tensors & Pseudo-Tensors \\ \hline \noalign{\medskip}
		1 & $\frac{\hbar}{r^2}$ & $V_0(r)$ &  &  & (\ref{Saxial})& \eqref{eq45}, \eqref{eq47}, \eqref{eq49} & $-$ \\\noalign{\medskip}
		& & $\frac{\hbar^2}{r^2}$ &  & (\ref{Piaxial}) & (\ref{Saxial}) & \eqref{eq45}, \eqref{eq47}, \eqref{eq49} & \eqref{eqA6}$-$\eqref{eqA9} \\\noalign{\medskip}
		& & $\frac{\hbar^2}{r^2}+\alpha r^2$ &  &  &  & \eqref{eq45}$-$\eqref{eq49} & $-$ \\\noalign{\medskip}
		& & $\frac{\hbar^2}{r^2}-\frac{\alpha}{r}$ &  &  &  & $-$ &  \eqref{eqA10} \\\noalign{\medskip}
		2 & $\frac{\hbar}{2r^2}$ & $V_0(r)$ & (\ref{pscase1int1}), (\ref{pscase1int2}) & (\ref{firstcasesub11finint3}), (\ref{firstcasesub11finint4}) & $-$ & $-$ & \eqref{eqY6} \\\noalign{\medskip}
		&  & $\frac{3\hbar^2}{8r^2}-\frac{\alpha}{r}$ & (\ref{pscase1int1}), (\ref{pscase1int2}) & (\ref{firstcasesub11finint1})$-$(\ref{firstcasesub11finint4}) & $-$  & $-$ & $-$ \\\noalign{\medskip}
		3 & $-\frac{\hbar}{2r^2}$ & $V_0(r)$ & $-$ & $-$ & $-$ & $-$ & \eqref{eqY7} \\\noalign{\medskip}
		&  & $-\frac{\hbar^2}{8r^2}+\alpha r^2$ & (\ref{intofmotforcase2}) & $-$ & $-$ & $-$ & $-$ \\\noalign{\medskip}
		4 & $\frac{3\hbar}{2r^2}$ & $V_0(r)$ & $-$ & $-$ & $-$ & $-$  & \eqref{eqY5} \\\noalign{\medskip}
		& & $\frac{15\hbar^2}{8r^2}+\alpha r^2$ & (\ref{intofmotforcase3}) & $-$ & $-$ & $-$  & $-$ \\\noalign{\medskip}
		5 & $-\frac{\hbar}{r^2}$ & $\alpha$ & $-$ & $-$ & $-$  & $-$ & \eqref{eqY3}, \eqref{eqY4} \\\noalign{\medskip}
		&   & $\alpha r^2$ & $-$ & $-$ & (\ref{axialveccase23})  & $-$ & $-$ \\\noalign{\medskip}
		6 & $\frac{2\hbar}{r^2}$ & $\frac{3\hbar^2}{r^2}$ & $-$ & $-$ & $-$ & $-$ & \eqref{eqY1}, \eqref{eqY2} \\\noalign{\medskip}
		&  & $\frac{3\hbar^2}{r^2}+\alpha r^2$ & $-$ & $-$ & (\ref{axialveccase30}) & $-$ & $-$ \\\noalign{\medskip}
		7 & $\frac{\alpha}{2r^2}+\beta$ & $\frac{\alpha(\alpha+2\hbar)+4\beta r^4}{8r^2}$ & $-$ & (\ref{finintofmotionveccase1a}), (\ref{finintofmotionveccase1}) & $-$  & $-$ & $-$ \\\noalign{\medskip}
		8 & $\frac{\hbar}{2r^2}\left( 1+ \frac{\epsilon}{\sqrt{1+\beta r^2}}\right)$ & $\frac{\hbar^2}{2r^2}\left( 1+ \frac{\epsilon}{\sqrt{1+\beta r^2}}\right)$ & (\ref{case4intofmotfromoldart}), (\ref{case4intofmot}) & (\ref{revisedadditionaliomvector}) & $-$ & $-$ & $-$ \\\noalign{\medskip}
		9 & $\frac{\hbar}{2r^2}\left( 1+ \frac{2 \epsilon}{\sqrt{1+\beta r^2}}\right)$ & (\ref{case5f2}) & (\ref{case5intofmot}) & $-$ & $-$ & $-$ &  $-$ \\\noalign{\medskip}
		10 & $\frac{\epsilon\hbar}{r^2\sqrt{1+\beta r^2}}$ & (\ref{axialcase4V0sol}) & $-$ & $-$ & (\ref{newaxialcase4intofmot3}) & $-$ & $-$ \\\noalign{\medskip}
		11 & $\frac{\hbar}{r^2}\left( 1+ \frac{\epsilon}{\sqrt{1+\beta r^2}}\right)$ & (\ref{axialcase5V0sol}) & $-$ & $-$ & (\ref{newaxialcase5intofmot3}) & $-$ & $-$ \\
		\hline
	\end{tabular}
\end{table}
\end{landscape}
The scalars potentials that do not fit into the table are given below:
\begin{equation}
	V_0 = \frac{\hbar^2}{8 r^2\left(1+\beta r^2\right)^2}\left(7+10 r^2 \beta +8\epsilon \left(1+\beta r^2\right)^{3/2}\right) - \frac{\alpha}{4\beta(1+\beta\,r^2)}\,,
	\label{case5f2}
\end{equation}
\begin{eqnarray}
	V_0 = \frac{\hbar^2}{8 r^2 \left(1+\beta r^2\right)^2} \left( 4+6 \beta r^2 - r^4 \beta ^2 + 4 \epsilon\, \left(1+\beta r^2 \right)^{\frac{3}{2}}\right)+\frac{\alpha }{1+\beta r^2}\,,
	\label{axialcase4V0sol}
\end{eqnarray}
\begin{eqnarray}
	V_0 = \frac{3\hbar^2 \Big(4+5 \beta r^2+4\epsilon \left(1 + \beta r^2\right)^{\frac{3}{2}}\Big)}{8 r^2 \left(1+\beta r^2\right)^2}-\frac{\alpha }{2 \beta \left(1+\beta r^2\right)}\,.
	\label{axialcase5V0sol}
\end{eqnarray}
\section{Acknowledgments}
\.{I}Y thanks the Centre de Recherches Math\'{e}matiques, Universit\'{e} de Montr\'{e}al, 
where part of this work was done, for the kind hospitality.

This research program, systematic investigation of superintegrability with spin, was initiated by Pavel Winternitz in 2005. Throughout the program he was the coauthor 
of the most of the articles. This last work started in 2018 when \.{I}Y was visiting the Centre de Recherches Math\'{e}matiques, Universit\'{e} de Montr\'{e}al on a sabbatical leave. 
Pavel Winternitz was involved in almost every stage of the article and was eager to see its completion. However, he passed away on February 13, 2021.  We dedicate this paper to his memory.

\newpage
\appendix
\section{Integrals of Motion for the Gauge Induced Potential}
In the previous sections, it has been pointed out that we excluded the special case $V_1=\frac{\hbar}{r^2}$ since it is a gauge induced potential. Nevertheless, we present the (pseudo)tensor integrals of motion appeared in such cases.  

For the gauge induced potential $V_1=\frac{\hbar}{r^2}$, we have the following five tensor integrals of motion:
\begin{align} \allowdisplaybreaks
\mathcal{T}_2^{\,ij}=&(L^iL^j+L^jL^i)+2\hbar(L^i\sigma^j+\sigma^iL^j)+\dfrac{2\hbar}{r^2}\big( 2(\vec{\sigma},\vec{L})+\hbar\big)x^ix^j \nonumber
\\&-2\hbar\big( p^i(\vec{x}\wedge\vec{\sigma})^j+(\vec{x}\wedge\vec{\sigma})^i p^j\big)+\dfrac{2\hbar}{r^2}\big( (\vec{x},\vec{p})+\mathrm{i}\hbar\big)\big(x^i(\vec{x}\wedge\vec{\sigma})^j+(\vec{x}\wedge\vec{\sigma})^ix^j\big),\label{eq45}\\
\mathcal{T}_3^{\,ij}=&2(p^ip^j+2\alpha\, x^ix^j)-\dfrac{2\hbar}{r^2}\pig( \dfrac{\hbar}{r^2}x^ix^j+\big(p^i(\vec{x}\wedge\vec{\sigma})^j+(\vec{x}\wedge\vec{\sigma})^ip^j\nonumber\\&+\dfrac{\mathrm{i}\hbar}{r^2}( x^i(\vec{x}\wedge\vec{\sigma})^j+(\vec{x}\wedge\vec{\sigma})^ix^j ) \big) \pig),\label{eq46}\\
\mathcal{T}_4^{\,ij}=&\mathrm{i}\hbar(L^i\sigma^j+\sigma^iL^j)-\big( 2(\vec{\sigma},\vec{L})+\hbar \big)(x^ip^j+p^ix^j) -(\vec{x},\vec{p})(x^i(\vec{p}\wedge\vec{\sigma})^j+(\vec{p}\wedge\vec{\sigma})^ix^j) \nonumber\\
&+\dfrac{2}{r^2}\big( \mathrm{i}\hbar^2+\hbar(\vec{x},\vec{p})+2\big((\vec{x},\vec{p})+\mathrm{i}\hbar \big)(\vec{\sigma},\vec{L})\big)x^ix^j+r^2(p^i(\vec{p}\wedge\vec{\sigma})^j+(\vec{p}\wedge\vec{\sigma})^ip^j)\nonumber\\
&-\big( (\vec{x},\vec{p})+2\,\mathrm{i}\hbar \big)(p^i(\vec{x}\wedge\vec{\sigma})^j+(\vec{x}\wedge\vec{\sigma})^ip^j) \nonumber\\
&+\big(-\dfrac{2}{r^2}\vec{L}^{\:2}+\vec{p}^{\:2}-\dfrac{3\hbar^2}{2r^2} \big)(x^i(\vec{x}\wedge\vec{\sigma})^j+(\vec{x}\wedge\vec{\sigma})^ix^j),\label{eq47}\\
\mathcal{T}_5^{\,ij}=&\dfrac{(3\hbar+2(\vec{\sigma},\vec{L}))}{r^2}\pig(\dfrac{2\,\mathrm{i}\hbar}{r^2}x^ix^j+(x^ip^j+p^ix^j)\pig)-(p^i(\vec{p}\wedge\vec{\sigma})^j+(\vec{p}\wedge\vec{\sigma})^ip^j)\nonumber\\
&+\dfrac{2}{r^2}\big( (\vec{x},\vec{p})+\mathrm{i}\hbar \big) (p^i(\vec{x}\wedge\vec{\sigma})^j+(\vec{x}\wedge\vec{\sigma})^ip^j)\nonumber\\
&+\big( 2\alpha-\dfrac{\hbar}{r^4}(3\hbar-2\,\mathrm{i}(\vec{x},\vec{p})) \big)(x^i(\vec{x}\wedge\vec{\sigma})^j+(\vec{x}\wedge\vec{\sigma})^ix^j), \label{eq48}\\
\mathcal{T}_6^{\,ij}=&-(L^i\sigma^j+\sigma^iL^j)-\dfrac{2}{r^2}\big( 2(\vec{\sigma},\vec{L})+\hbar \big)x^ix^j+2(p^i(\vec{x}\wedge\vec{\sigma})^j+(\vec{x}\wedge\vec{\sigma})^ip^j)\nonumber\\
&-\dfrac{2}{r^2}\big( (\vec{x},\vec{p})+\mathrm{i}\hbar\big)(x^i(\vec{x}\wedge\vec{\sigma})^j+(\vec{x}\wedge\vec{\sigma})^ix^j).\label{eq49}
\end{align}
We observe two scalar potentials for the gauge induced potential which are $V_0=\frac{\hbar^2}{r^2}$ and $V_0=\frac{\hbar^2}{r^2}+\alpha r^2$. We see that the integrals \eqref{eq45}, \eqref{eq47} and \eqref{eq49} exist for any $V_0(r)$ or $V_0=\frac{\hbar^2}{r^2}$. On the other hand, \eqref{eq46} and \eqref{eq48} exist only for $V_0=\frac{\hbar^2}{r^2}+\alpha r^2$.  

For the gauge induced potential $V_1=\frac{\hbar}{r^2}$, we have the following pseudo-tensor integrals of motion:
\begin{align}
\mathcal{Y}_8^{\,ij}=&(p^iL^j+L^ip^j)-\dfrac{\hbar}{r^2}\big( (\vec{x},\vec{p})+\mathrm{i}\hbar \big)(x^i\sigma^j+\sigma^ix^j)+\dfrac{\hbar}{r^2}(\vec{\sigma},\vec{x})(x^ip^j+p^ix^j)\nonumber\\
&+\dfrac{2\hbar}{r^2}\big( (\vec{\sigma},\vec{p})+\dfrac{2\mathrm{i}\hbar}{r^2}(\vec{\sigma},\vec{x}) \big)x^ix^j+\dfrac{\hbar}{2}(p^i\sigma^j+\sigma^i p^j),\label{eqA6}\\
\mathcal{Y}_9^{\,ij}=&\dfrac{2\mathrm{i}\hbar}{r^2}\big((\vec{\sigma},\vec{x})(x^ip^j+p^ix^j)-2(\vec{\sigma},\vec{p})x^ix^j-r^2(p^i\sigma^j+\sigma^ip^j)\big)\nonumber\\
&-\big((\vec{p}\wedge\vec{\sigma})^iL^j+L^i(\vec{p}\wedge\vec{\sigma})^j \big)+\dfrac{2}{r^2}(\vec{x},\vec{p})\big((\vec{x}\wedge\vec{\sigma})^iL^j+L^i(\vec{x}\wedge\vec{\sigma})^j \big)\nonumber\\
&+\dfrac{\hbar}{r^2}\big( 2\mathrm{i}(\vec{x},\vec{p})-\dfrac{\hbar}{2}\big)(x^i\sigma^j+\sigma^ix^j)+\dfrac{1}{r^2}\big( 2(\vec{\sigma},\vec{L})+3\hbar \big)(x^iL^j+L^ix^j),\\
\mathcal{Y}_{10}^{\,ij}=&-\dfrac{\hbar}{r^2}(x^iL^j+L^ix^j)-\big((\vec{p}\wedge\vec{\sigma})^iL^j+L^i(\vec{p}\wedge\vec{\sigma})^j \big)-(\vec{x},\vec{p})(p^i\sigma^j+\sigma^ip^j)\nonumber\\
&-\dfrac{2}{r^2}(\vec{\sigma},\vec{x})\big( (\vec{x},\vec{p})+\mathrm{i}\hbar \big)(x^ip^j+p^ix^j)+\vec{p}^{\:2}(x^i\sigma^j+\sigma^ix^j)\nonumber
\\ &+4(\vec{\sigma},\vec{x})p^ip^j+\dfrac{4\hbar}{r^4}(\vec{\sigma},\vec{x})\big(\hbar-\mathrm{i}(\vec{x},\vec{p}) \big)x^ix^j,\\
\mathcal{Y}_{11}^{\,ij}=&(p^i\sigma^j+\sigma^ip^j)-\dfrac{2}{r^2}(\vec{\sigma},\vec{x})(x^ip^j+p^ix^j)-\dfrac{4\mathrm{i}\hbar}{r^4}(\vec{\sigma},\vec{x})x^ix^j,
\label{eqA9}\\ 
\mathcal{Y}_{12}^{\,ij}=&\dfrac{2}{r^2}\bigg( -2\mathrm{i}\hbar\big((\vec{\sigma},\vec{p})+\dfrac{1}{r^2}(\vec{x},\vec{p}) (\vec{\sigma},\vec{x})\big)+\big(\dfrac{\alpha}{r}+\dfrac{\hbar^2}{r^2}\big)(\vec{\sigma},\vec{x}) \bigg)x^i x^j\nonumber\\&-\big( 2\mathrm{i}\hbar+(\vec{x},\vec{p}) \big)(p^i\sigma^j+\sigma^ip^j)-\dfrac{2}{r^2}(\vec{x},\vec{p})(\vec{\sigma},\vec{x})(x^ip^j+p^ix^j)+4(\vec{\sigma},\vec{x})p^ip^j\nonumber\\
&-2\big((\vec{p}\wedge\vec{\sigma})^iL^j+L^i(\vec{p}\wedge\vec{\sigma})^j \big)+\dfrac{2}{r^2}(\vec{x},\vec{p})\big((\vec{x}\wedge\vec{\sigma})^iL^j+L^i(\vec{x}\wedge\vec{\sigma})^j \big)\nonumber\\
&+\big(\dfrac{2\mathrm{i}\hbar}{r^2}(\vec{x},\vec{p})+\vec{p}^{\:2}-(\dfrac{\alpha}{r}+\dfrac{3\hbar^2}{r^2})\big)(x^i\sigma^j+\sigma^ix^j)+\dfrac{2}{r^2}\big( (\vec{\sigma},\vec{L})+\hbar \big)(x^iL^j+L^ix^j)\,. \label{eqA10}
\end{align}
The last pseudo-tensor integral of motion can also be represented as 
\begin{align}
\mathcal{Y}_{12}^{\,ij}
=& \mathcal{Y}_{9}^{\,ij}+\mathcal{Y}_{10}^{\,ij}+\dfrac{\alpha}{r}\big( \dfrac{2}{r^2}(\vec{\sigma},\vec{x})x^ix^j-(x^i\sigma^j+\sigma^ix^j)  \big)   \label{eqA11}.
\end{align}
There are two scalar potentials for this case that are $V_0=\frac{\hbar^2}{r^2}$ and $V_0=\frac{\hbar^2}{r^2}-\frac{\alpha}{r}$. The integrals \eqref{eqA6}--\eqref{eqA9} exist for $V_0=\frac{\hbar^2}{r^2}$ whereas the other one \eqref{eqA10} exists for the potential $V_0=\frac{\hbar^2}{r^2}-\frac{\alpha}{r}$.
\section{Pseudo-scalar, Vector and Axial Vector Integrals of Motion}
In this appendix, we remind the pseudo-scalar, vector and axial vector integrals of motion presented in \cite{DWY}. No nontrivial scalar integrals of motion exist.

\noindent\textbf{Pseudo-scalars}:
\begin{align}
		&X_P^1 = \frac{(\vec{\sigma}, \vec{x})}{r}\,, \label{pscase1int1} \\
	&X_P^2 = -r (\vec{\sigma}, \vec{p}) + \frac{1}{r}\,(\vec{\sigma}, \vec{x})\,(\vec{x}, \vec{p}) - \frac{i\hbar}{r}\,(\vec{\sigma}, \vec{x})\,,
	\label{pscase1int2}\\
	&	X_P^3 =  \frac{(\vec{\sigma}, \vec{x})}{r}\, \left(\frac{3\hbar^2}{2\,r^2} + 4\alpha\,r^2 - 2 (\vec{p},\vec{p}) \right) + 
	\frac{4}{r}\,\big((\vec{x}, \vec{p}) - i\hbar\big)\, (\vec{\sigma}, \vec{p})\,,
	\label{intofmotforcase2}\\
		&X_P^4 =  \frac{(\vec{\sigma}, \vec{x})}{r}\, \left(-\frac{5\hbar^2}{2\,r^2} + 4\,\alpha\,r^2 - \frac{20i\hbar}{r^2}\,(\vec{x},\vec{p}) - 2 (\vec{p},\vec{p}) + \frac{8}{r^2} \big(\vec{x}, (\vec{x}, \vec{p})\,\vec{p}\big) \right) - 
	\frac{4}{r}\,\big((\vec{x}, \vec{p}) - 2i\hbar\big)\, (\vec{\sigma}, \vec{p})\,,
	\label{intofmotforcase3}\\
	&	X_P^5 = -\frac{1}{\beta}\sqrt{1+\beta r^2}\,(\vec{\sigma}, \vec{p}) + 
	\frac{(\vec{\sigma}, \vec{x})}{-\epsilon + \sqrt{1+\beta r^2}}\,\,  \Big((\vec{x}, \vec{p}) -i\hbar \Big)\,,
	\label{case4intofmotfromoldart}\\
	&	X_P^6 = \frac{2}{r^2}\big(1+\epsilon\, \sqrt{1+\beta\,r^2}\big)\,(\vec{\sigma}, \vec{x})\, 
	\Big(\hbar^2 + 3 i\hbar \,(\vec{x}, \vec{p}) - \big(\vec{x}, (\vec{x}, \vec{p})\,\vec{p}\big)\Big) \nonumber \\ 
	&\qquad\quad- 2i\hbar \Big(2+\epsilon\, \sqrt{1+\beta\,r^2}\Big)\, (\vec{\sigma}, \vec{p}) 
	+ 2\,(\vec{x}, \vec{p})\,(\vec{\sigma}, \vec{p}) + 2 \epsilon\, \sqrt{1+\beta\,r^2}\,(\vec{\sigma}, \vec{x})\,(\vec{p}, \vec{p})\,,
	\label{case4intofmot}\\
	&	X_P^7 = \frac{2\,r^4\,\alpha\,(1 + \beta\,r^2) - \hbar^2(1 + 2\beta\,r^2)\,\Big(1 + 4\epsilon\sqrt{1+\beta\,r^2} + 4r^2\beta\,\big(1 +\epsilon\, \sqrt{1 + \beta\,r^2}\big)\Big)}{2\,r^3\,(1 + \beta\,r^2)^2} \,(\vec{\sigma}, \vec{x}) \nonumber \\
	&\qquad\quad- \frac{2i\hbar}{r^3}\left(5 + 3 \beta\, r^2 - \frac{\epsilon}{\sqrt{1+\beta\,r^2}} + 6\,\epsilon\, \sqrt{1+\beta\,r^2}\right)\,(\vec{\sigma}, \vec{x})\,(\vec{x}, \vec{p}) \nonumber \\
	&\qquad\quad+ \frac{2i\hbar}{r}\left(1 + \beta\, r^2 - \frac{\epsilon}{\sqrt{1+\beta\,r^2}} + 4\,\epsilon\, \sqrt{1+\beta\,r^2}\right)\,(\vec{\sigma}, \vec{p}) \nonumber \\
	&\qquad\quad- \frac{2}{r}\,(1 + \beta\,r^2)\,(\vec{\sigma}, \vec{x})\,(\vec{p}, \vec{p}) + \frac{2}{r^3}\, \Big(2 + \beta\, r^2 + 2\,\epsilon\, \sqrt{1+\beta\,r^2} \Big)\,(\vec{\sigma}, \vec{x})\,\big(\vec{x},(\vec{x}, \vec{p})\vec{p} \big) \nonumber \\
	&\qquad\quad- \frac{4\epsilon}{r}\,\sqrt{1 + \beta\,r^2}\,(\vec{x}, \vec{p})\,(\vec{\sigma}, \vec{p})\,.
	\label{case5intofmot}
\end{align}
\textbf{Vectors}:
\begin{align}
	&\vec{X}_V^1 = - \big(\vec{\sigma}, \vec{L}\big)\,\vec{p} + \frac{3\hbar}{2}\vec{p} - \frac{\hbar\vec{x}}{2 r^2} (\vec{x}, \vec{p}) + \frac{i\hbar}{2}(\vec{\sigma} \wedge \vec{p}) + i\hbar^2 \frac{\vec{x}}{2 r^2} + \frac{4 \alpha r - \hbar^2}{4 r^2} (\vec{x} \wedge \vec{\sigma})\,,
	\label{firstcasesub11finint1} \\
	&\vec{X}_V^2 = 2 \vec{x} \vec{p}^{\,\,2} - 2 (\vec{x}, \vec{p})\,\vec{p} - \frac{\hbar\vec{x}}{r^2} \big(\vec{\sigma}, \vec{L}\big) + 2 i\hbar \vec{p} - \hbar(\vec{\sigma} \wedge \vec{p}) + i\hbar^2 \frac{(\vec{x} \wedge \vec{\sigma})}{2 r^2} + \frac{\vec{x}}{r^2} (\hbar^2-2\alpha r)\,,
	\label{firstcasesub11finint2} \\ 
	&\vec{X}_V^3 = \frac{(\vec{x}, \vec{\sigma})}{r} \vec{L} + \frac{\hbar}{2 r} \big(\vec{x} - i(\vec{x} \wedge \vec{\sigma})\big)\,,
	\label{firstcasesub11finint3} \\
	&\vec{X}_V^4 = r \vec{L} (\vec{\sigma},\vec{p}) + \frac{\hbar r}{2}\vec{p} - \frac{i\hbar r}{2} (\vec{\sigma} \wedge \vec{p}) + \vec{X}_V^3\,\Big(i\hbar  - (\vec{x}, \vec{p})\Big)\,,
	\label{firstcasesub11finint4}\\
		\label{finintofmotionveccase1a}
	&\vec{X}_V^5 = \vec{x}\,\Big(2\vec{p}^{\,2} - \left(\frac{2\beta r^2 -\alpha}{r^2}\right) \Big(\big(\vec{\sigma}, \vec{L}\big) + \hbar\Big) \Big) 
	+ 2 \big(i\hbar - (\vec{x}, \vec{p})\big)\,\vec{p} - \hbar(\vec{\sigma} \wedge \vec{p}) \nonumber \\
	&\qquad \,\,+ i\hbar \left(\frac{2\beta r^2 -\alpha}{2 r^2}\right)\,(\vec{\sigma} \wedge \vec{x})\,,  \\
	&\vec{X}_V^6 = \Big(\frac{1}{2} (2 \hbar +\alpha -2 \beta r^2) + \big(\vec{\sigma}, \vec{L}\big) \Big)\,\vec{p} + \frac{\vec{x}}{2}\left(\frac{2\beta r^2 -\alpha}{r^2}\right) \Big((\vec{x}, \vec{p}) - i\hbar\Big) + \frac{i\hbar}{2} (\vec{\sigma} \wedge \vec{p}) \nonumber \\
	& \qquad\quad - \frac{\hbar}{4}\left(\frac{2\beta r^2 -\alpha}{r^2}\right) (\vec{\sigma} \wedge \vec{x})\,, 
	\label{finintofmotionveccase1}\\
	&\vec{X}_V^7 = \{X_P^1, \vec{\mathcal{J}}\}\,.
	\label{revisedadditionaliomvector}
\end{align}
\textbf{Axial vectors:}
\begin{align}
		&\vec{X}_A^1 =  -\big(2\alpha r^2 + \vec{p}^{\,\,2}\big)\vec{\sigma} + 2 (\vec{\sigma},\vec{p}) \vec{p} \nonumber \\
	&\qquad\quad + \frac{2}{r^2}\Bigg(\vec{x} \Big((\vec{\sigma},\vec{x})\,\vec{p}^{\,\,2} + 2 i\hbar (\vec{\sigma},\vec{p}) - 2 (\vec{x},\vec{p})(\vec{\sigma},\vec{p})\Big) + i\hbar \big(\vec{\sigma} (\vec{x},\vec{p}) - (\vec{\sigma},\vec{x}) \vec{p}\,\big)\Bigg)\,,
	\label{axialveccase23}\\	
	&\vec{X}_A^2 = \big(3 \vec{p}^{\,\,2} - 2\alpha r^2+\frac{4}{r^2}\left(i\hbar (\vec{x},\vec{p})-(\vec{x},(\vec{x},\vec{p}) \vec{p})\right)\big)\vec{\sigma} -\frac{2}{r^2}\big(\hbar+(\vec{L},\vec{\sigma})\big)\vec{L} -2\left((\vec{\sigma},\vec{p})-\frac{3i \hbar(\vec{\sigma},\vec{x})}{r^2}\right)\vec{p} \nonumber \\ 
	&\qquad\quad + \frac{2 \vec{x}}{r^4}\bigg(3i\hbar r^2(\vec{\sigma},\vec{p})-(\vec{\sigma},\vec{x})\big(3h^2-2r^4\alpha+2r^2\vec{p}^{\,2}+12i\hbar (\vec{x},\vec{p})\big)-4(\vec{x},(\vec{x},\vec{p}) \vec{p})\bigg) \,,
	\label{axialveccase30}\\	
		&\vec{X}_A^3 = \left( 2 i\hbar \frac{(\vec{\sigma}, \vec{x})}{r^2} Q_+  - 4 \epsilon \sqrt{1+\beta r^2} \, (\vec{\sigma},\vec{p})  \right) \vec{p} - \frac{4}{r^2}\left(\hbar + 2 \alpha r^2 + \frac{\epsilon\hbar}{\sqrt{1+\beta r^2}} + q(\vec{\sigma},\vec{L})\right)\,  \vec{L} \nonumber \\
	&\qquad\quad +  \frac{\vec{\sigma}}{2 r^2} \Bigg(Y + 8 i\hbar \left(1+\frac{\beta}{2} r^2\right)(\vec{x},\vec{p}) - 8 q (\vec{x},(\vec{x},\vec{p}) \vec{p}) + 4 r^2 \left(2q-1\right)\vec{p}^{\,\,2} \Bigg) 
	\nonumber \\ 
	&\qquad\quad +  \frac{2 \vec{x}}{r^2} \Bigg(\bigg( i\hbar Q_- + 4\epsilon \sqrt{1+\beta r^2} (\vec{x},\vec{p})\bigg) (\vec{\sigma},\vec{p}) - 2\epsilon (\vec{\sigma},\vec{x}) \sqrt{1+\beta r^2} \vec{p}^{\,\,2} \Bigg)\,,
	\label{newaxialcase4intofmot3}\\	
		&\vec{X}_A^4 = \Bigg( \frac{2 i\hbar^2 (\vec{\sigma} , \vec{x})}{r^2} \widetilde{Q} - 4 \epsilon\, \sqrt{1+\beta r^2}\, (\vec{\sigma},\vec{p})  \Bigg) \vec{p} - \frac{4}{r^2}\Bigg(\hbar + \frac{\epsilon\hbar}{\sqrt{1 + \beta r^2}} + q (\vec{\sigma}, \vec{L}) \Bigg) \vec{L} \nonumber \\
	&\qquad\quad +  \frac{2 \vec{\sigma}}{r^2} \Bigg(2 i\hbar q (\vec{x},\vec{p}) - \widetilde{Y}  - 2 q (\vec{x},(\vec{x},\vec{p}) \vec{p}) + r^2 \left(2q-1\right)\vec{p}^{\,\,2} \Bigg) 
	+ \frac{2 \vec{x}}{r^4} \Bigg((\vec{\sigma},\vec{x}) \Big(4 q (\vec{x},(\vec{x},\vec{p}) \vec{p}) \nonumber \\
	&\qquad\quad -  2 r^2 \left(2q-1-\epsilon\, \sqrt{1+\beta r^2}\right) \vec{p}^{\,\,2} -
	Z - 4 i\hbar W (\vec{x},\vec{p})\Big) + i\hbar \widetilde{Q}(\vec{\sigma},\vec{p})  \Bigg)\,,
	\label{newaxialcase5intofmot3}
\end{align}
where $Q_{\pm}$, $q$, $Y$, $\widetilde{Q}$, $\widetilde{Y}$, 
$Z$ and $W$ are given by the following relations
\begin{align*}
		&Q_{\pm} = 1 + \frac{\beta}{2} r^2 \pm \epsilon\frac{3 + 4 \beta r^2}{\sqrt{1+\beta r^2}}\,, \qquad 
	q = 1+\frac{\beta}{2} r^2+\epsilon\, \sqrt{1+\beta r^2}\,, \\
	&Y = \frac{-4 \hbar^2 - 8 r^2 \alpha - 6\hbar^2 r^2 \beta - 8r^4\alpha \beta + \hbar^2 r^4 \beta^2}{(1 + \beta r^2)^2} - \frac{4\epsilon \hbar^2}{\sqrt{1+\beta r^2}}\,, \\
	&	\widetilde{Q} = 3 + \frac{5 \beta}{2} r^2 + \epsilon\frac{3 + 4 \beta r^2}{\sqrt{1+\beta r^2}}\,, \qquad 
	W = 3+ 2 \beta r^2 + \epsilon\, \frac{6 + 7 \beta r^2}{2 \sqrt{1+\beta r^2}}\,, \\
	&\widetilde{Y} = \frac{\hbar^2 r^4 (4 \alpha - 6 \beta^2) + \hbar^2 r^6(3 \beta^3 + 4\beta \alpha) -6 \hbar^2 \beta r^2}{4 (1 + \beta r^2)^2} + \frac{2\epsilon \hbar^2 \beta r^2}{\sqrt{1+\beta r^2}}\,, \\
&	Z = \frac{-4 r^4 \alpha (1 + \beta r^2) + 3 \hbar^2 \big(2 + \beta r^2 (7 + 6 \beta r^2)\big)}{2 (1 + \beta r^2)^2} + \epsilon \frac{3 \hbar^2 (1 + 2 \beta r^2)}{\sqrt{1 + \beta r^2}}\,.
\end{align*}


\end{document}